解説・講座・小特集

# Burst Intensification by Singularity Emitting Radiation in Laser Plasma

PIROZHKOV Alexander S.[1)*], ESIRKEPOV Timur Zh.[1)*], PIKUZ Tatiana A.[2)], SAGISAKA Akito[1)], OGURA Koichi[1)], KOGA James K.[1),3)], BIERWAGE Andreas[4)], KIRIYAMA Hiromitsu[1)], NAMBA Shinichi[5)], BULANOV Sergei V.[1),6)], KANDO Masaki[1)]

National Institutes for Quantum Science and Technology (QST), Kansai Institute for Photon Science (KPSI)[1)],

Osaka University, Institute for Open and Transdisciplinary Research Initiatives[2)],

Kyoto International University Academy[3)]

National Institutes for Quantum Science and Technology (QST), Rokkasho Institute for Fusion Energy[4)]

Hiroshima University, Graduate School of Advanced Science and Engineering[5)]

Extreme Light Infrastructure ERIC, ELI Beamlines Facility[6)]

[*] Equally-contributing authors

**Abstract:** Burst Intensification by Singularity Emitting Radiation (BISER) appears as a bright temporally and spatially coherent Extreme Ultraviolet (XUV) and x-ray source driven by compact multi-terawatt femtosecond lasers in gas targets. There BISER originates from relativistic plasma singularities, so that the emission source size has a nanometer scale. The BISER x-ray yield quadratically depends on the driving laser power. BISER spectra have hundreds of electronvolt (eV) bandwidth embracing the 'water window' region (284 – 543 eV). Simulations predict that BISER pulses have durations close to the transform limit, which promises pulses shorter than the atomic unit of time (24 attoseconds). Based on the BISER brightness at ~20 terawatt laser power and the quadratic scaling, the brightness of BISER driven by petawatt-class lasers is predicted to exceed XUV free electron lasers. The BISER concept creates a new framework for a wide range of media emitting travelling waves capable of constructive interference, including gravitational waves. Here we review the BISER experimental discovery, its explanation based on the relativistic laser plasma simulations and catastrophe theory, experimental validation of the theoretical BISER model, proposal for imaging fast moving singularities, driving laser requirements, and future prospects emphasizing the driving laser wavelength scalability and possibility of terawatt attosecond coherent x-ray pulse generation.

## 1. Introduction

Coherent x-ray sources are indispensable in fundamental research and applications. Recent trends in development of such sources include two broad classes: kilometer-scale accelerator-based x-ray FELs [1] and relatively compact (a few meters long) laser-based sources such as x-ray lasers [2] and atomic high-order harmonics [3]. Fundamental limitations of conventional techniques related to the above-mentioned classes severely hinder the development of a bright compact coherent x-ray source, especially at keV photon energies [2, 4, 5]. Relativistic laser plasma [6] brings new classes of bright coherent sources [7], based on high frequency radiation generation (including high-order harmonic generation [8]) due to nonlinearities or/and Doppler effect peculiar to relativistic dynamics. Some examples of such sources are the Akhiezer-Polovin mechanism of high-order harmonic generation [9], nonlinear Thomson scattering [10-13], electromagnetic soliton synchrotron-like afterglow [14], the Relativistic Oscillating Mirror (ROM) [15], the Relativistic Sliding Mirror (RSM) [16, 17], the Relativistic Flying Mirror (RFM) [18], the relativistic forcibly oscillating diffractive grating [19], and high-frequency radiation from relativistic quasiparticles [20].

Here we review the recent development of the concept of Burst Intensification by Singularity Emitting Radiation (BISER). We discovered BISER in the interaction of a femtosecond multi-TW laser pulse with a gas jet [21, 22]. BISER appears as ultra-bright spatially and temporally coherent XUV and soft x-ray radiation with a yield proportional to the laser power squared [22], source size in the nanometer range [23], and attosecond duration close to the transform limit [23].

The BISER mechanism is based on a constructive interference of travelling waves from emitting matter extremely concentrated in converging flows, Fig. 1. In a simplest model, two colliding flows in a medium consisting of elementary emitters create a density spike. What does a distant observer detecting the travelling waves see? Assume that the density spike is sufficiently high, so that in its vicinity, for a finite emitted wavelength $\lambda_E$ there is a region $\mathcal{R}_{\lambda_E}$ where the distance between the elementary emitters is much shorter than $\lambda_E$. In the "worst" case scenario of a random phase of elementary emitters, the distant observer may see a burst of *spatially coherent* emission from the region $\mathcal{R}_{\lambda_E}$ around the emerging spike, with an intensity proportional to the elementary emitter number in this region, $I_{\lambda_E} \propto N(\mathcal{R}_{\lambda_E})$. The most interesting case is when the elementary emitters have (nearly) the same phase. This condition is satisfied, e.g., when the elementary emitters are driven (excited) by an external field, whose spatial scale is much greater than $\lambda_E$, so that the emitters' phase slowly varies in space. In this case, the waves emitted from the region $\mathcal{R}_{\lambda_E}$ interfere constructively, so that the observer sees a burst of *spatially and temporally coherent* emission from the region $\mathcal{R}_{\lambda_E}$ around the emerging spike, with an intensity proportional to the square of the elementary emitter number in this region, $I_{\lambda_E} \propto N^2(\mathcal{R}_{\lambda_E})$. This conclusion can be partially reversed: if an observer detects an intense burst of *spatially and temporally coherent* emission with a *broad spectrum* in a relatively *wide solid angle* from an apparently small source, then the simplest hypothesis about the source size is the statement that this size is not greater than the emitted wavelength. Moreover, since the constructive interference occurs for any of the wavelengths in the assumed broad spectrum, the source is likely a sort of density spike where the emission source size is proportional to the emitted wavelength.

The formation of density spikes in a smoothly evolving physical system may seem unlikely or require very special conditions. Actually, density spikes naturally emerge in converging flows. Such flows ubiquitously occur in nature, e. g. with shock waves [24, 25] and jets [26-29] in astrophysical and laboratory plasmas [30-35] and, of course, in laser plasma [36, 37]. The emerging density spikes can be relatively long-lived and fast-moving. In the approximation of a continuous medium, which obeys the continuity equation (relating the density and velocity divergence), the density spike can be a singularity, where the density is formally infinite, but integrable – i.e., the singularity contains a finite mass. The emergence of a singularity can be regarded as a catastrophic change of density in the course of a smooth gradual evolution of the flow, which is the key subject of catastrophe theory [38, 39] focused on discontinuous changes of a system, despite continuous evolution of its control parameters. Catastrophe theory explains the guaranteed emergence of singularities in multi-stream flows, as well as their shape, universality and structural stability, i.e., an insensitivity of their existence to smooth perturbations. Physical media consist of discrete particles whose density is always finite. Nonetheless, the *density growth rate* near the peak of the density spike approximates that of the corresponding singularity in the continuous medium model; the denser the medium, the better is this approximation.

It is essential that the singularity consists of elementary emitters continuously flowing through it. In this respect it fundamentally differs from the case of a compact bunch, a fixed collection of elementary emitters moving together. In contrast to a compact bunch, the singularity has no restrictions on motion speed and acceleration. In particular, singularities can be superluminal, i.e. moving faster than the speed of light in vacuum [40, 41]. Further, the singularity has a non-local nature because it is determined by the global structure of the flow (e.g., inherits perturbations from distant parts of the flow). Nevertheless, in both cases – the singularity and a compact bunch, only the waves emitted from a sufficiently dense region can interfere constructively.

The BISER mechanism described above represents a fundamental general effect. Its formulation requires only the continuity equation (merely the conservation law of the particle number) for media capable of emitting traveling waves, e.g., electromagnetic or gravitational, and the general properties of an interference of travelling waves.

In this review, we investigate BISER in the case where the medium is a laser plasma and the travelling waves are electromagnetic ones. It appears that laser plasma is an ideal medium for investigating BISER in the laboratory, under controllable conditions. The converging flows, and, in general, multi-stream flows, naturally occur during propagation of a relativistically strong laser pulse in underdense plasma.

**Dimensionless parameters.** In the theoretical analysis and partly in the experimental planning and the description of the experimental results it is useful and often necessary to express all the physical variables via dimensionless parameters, which must be compared with dimensionless critical parameters indicating thresholds between different physical regimes. Here it is convenient to express distances and time in the laser wavelength $\lambda$ and period $T$, respectively; $T = 2\pi/\omega = \lambda/c$, where $\omega$ is the angular frequency of the laser pulse.

With respect to the electron dynamics, the laser field is characterized by the dimensionless amplitude, $a_0 = eE/(m_e c\omega)$, where $E$ is the electric field strength of the laser pulse, $e$ and $m_\mathrm{e}$ are the electron charge and mass, $c$ is the speed of light in vacuum. When $a_0 \gtrsim 1$, the electron dynamic is substantially relativistic [6], so the corresponding laser field is called relativistically strong. The laser irradiance (intensity) is $I = a_0^2 \times \pi c^5 m_\mathrm{e}^2/2e^2\lambda^2 = a_0^2 \times 1.37 \times 10^{18}\ \mathrm{W/cm^2} \times (1\ \mu\mathrm{m}/\lambda)^2$ (for a linearly polarized laser pulse). The laser pulse power is $\mathcal{P} = \iint I\, dydz = I_0 S_\mathrm{eff}$ (assuming the pulse propagates along the $x$-axis), and its energy is $\mathcal{E} = \int \mathcal{P}\, dt = P_0 \tau_\mathrm{eff}$; the integrals are determined by the shape of the laser pulse (in particular, the intensity distribution in the focal spot and temporal pulse profile). Here the formulae define the effective spot area, $S_\mathrm{eff} = \pi r_\mathrm{eff}^2$, spot radius, $r_\mathrm{eff}$, and pulse width (duration), $\tau_\mathrm{eff}$, via the integrals and the peak values of the irradiance, $I_0$, and power, $P_0$. The laser pulse is also characterized by the Full Width at Half Maximum (FWHM) spot size $D = D_\mathrm{FWHM}$ and pulse duration $\tau = \tau_\mathrm{FWHM}$, focal spot area $\mathcal{S} = \pi D^2/4$, F-number, etc. Laser pulse characteristics and their influence on BISER generation are discussed in Section **6**.

The electric field strength corresponding to the dimensionless amplitude of $a_0 = 1$ is $E_1 = 2\pi m_e c^2/e\lambda = \omega\sqrt{m_e/r_e} = 3.21 \times 10^{12}$ V/m × (1 μm/λ), where $r_e = e^2/m_e c^2 \approx 2.8 \times 10^{-13}$cm is the classical electron radius; for a typical laser wavelength of 1 μm, it is about 6 times greater than inter-atomic electric field of hydrogen, $E_a = \alpha^4\sqrt{m_e c^2/r_e^3} = 5.14 \times 10^{11}$ V/m, where $\alpha = e^2/\hbar c \approx 1/137$ is the fine-structure constant. Therefore, a relativistically strong laser pulse ionizes a low-Z gas via the optical field ionization (OFI) mechanism [42] in a small fraction of the laser period [43], producing fully ionized plasma (consisting of electrons and completely stripped ions). In such a case we can apply the approximation of a collisionless cold ideal plasma, which is described by the Vlasov equation and can be numerically simulated using the particle-in-cell (PIC) method [44]. The collisionless approximation also follows from a natural reduction of all collisional cross-sections at relativistic velocities [43].

The plasma density, $n_e$, is compared to the critical density, $n_{cr} = m_e\,\omega^2/(4\pi e^2) = \pi/(r_e\,\lambda^2) = 1.11 \times 10^{21}\text{cm}^{-3} \times (1\ \mu\text{m}/\lambda)^2$; plasma with $n_e/n_{cr} \ll 1$ is called underdense. The proper plasma oscillation is characterized by the Langmuir frequency $\omega_{pe} = \sqrt{4\pi e^2 n_e/m_e}$, $n_e/n_{cr} = \omega_{pe}^2/\omega^2$. Plasma is also characterized by the composition, ionization state of ions, etc.

In the following sections we show the results of the first experiments (Section **2**), their theoretical explanation including testable theoretical predictions (Section **3**), experimental confirmation of these predictions (Section **4**), a proposal of singularity observation (Section **5**), the experimentally deduced requirements of the quality of the laser system (Section **6**), a discussion of the obtainable BISER pulse duration (Section **7**), BISER driving laser wavelength scalability (Section **8**) and future applications (Section **9**); conclusions are in Section **10**.

## 2. Experimental discovery

Bright soft x-ray harmonic combs with highly unusual properties, later identified as BISER, were first discovered in an experiment [21] with the J-KAREN laser [45]. The experiment was repeated with the Astra Gemini laser [46], and the data from these two experiments were published together for the first time in the Letter [22]. Spatial coherence properties and off-axis emission of the new x-ray source were studied in Ref. [47]. A more detailed description of the available at the time results was provided in [48].

In the experiments, Fig. 2, multi-TW femtosecond lasers focused to irradiance of several units of $10^{18}$ W/cm$^2$ into high-purity helium gas jets with an atomic density of several units of $10^{19}$cm$^{-3}$, Table I, generated bright soft x-ray combs recorded by flat-field grazing-incidence spectrographs employing Varied Line Space (VLS) gratings with high resolving power [49-51]. The laser radiation, having multi-terawatt power even after transmission through the plasma, as well as low-frequency plasma radiation, were blocked by sub-micrometer-thick optical blocking filters [52, 53]. The high-frequency spectra transmitted through the filters extended up to the 'water window' spectral region (photon energy 284 – 543 eV) defined by the K absorption edges of carbon and oxygen; this spectral region is advantageous for imaging bio-samples in their natural wet and/or in-water state, as the water is transparent while the carbon-containing tissues are highly absorbing, thus enhancing the contrast.

**Table I. Parameters of lasers, gas targets, and plasma in the first experiments [21, 22, 47, 48]**

| Laser | J-KAREN | Astra Gemini |
|---|---|---|
| Pulse energy $E_L$ (J) | 0.4 | 3-10 |
| Pulse duration $\tau_{\text{FWHM}}$ (fs) | 27 | 54 |
| Peak power $P_0$ (TW) | 9 | 60 – 170 |
| Focusing OAP *f*-number | *f*/9 | *f*/20 |
| Focal spot FWHM (μm) | 10 | 25 × 40 |
| Focal spot at $1/e^2$ intensity (μm) | 20 | 40 × 80 |
| Vacuum irradiance (in vacuum, estimated) $I_{0,\text{Vac}}$ (W/cm$^2$) | $4 \times 10^{18}$ | $(2 – 5) \times 10^{18}$ |
| Dimensionless amplitude (in vacuum, estimated) $a_{0,\text{Vac}}$ | 1.4 | 1 – 1.5 |
| Gas target | Supersonic nozzle, He gas | |
| Nozzle orifice diameter (mm) | 1 | 0.5 |
| Nozzle Mach number | 3.3 | 2 |
| Nozzle – laser-axis distance (mm) | 1 | 1.5 |
| Peak electron density $n_{e,\text{max}}$ (cm$^{-3}$) | (1.5–9.6)×10$^{19}$ | (1.5–4.6)×10$^{19}$ |
| Amplitude after self-focusing (estimated) $a_{\text{SF}}$ | 5 – 8 | 12 – 16 |

High resolving power of the employed spectrographs allowed to resolve the periodic comb fringes, which evidenced the temporal coherence. The combs consisted of both odd and even orders of the base frequency, exhibited no difference in properties between the odd and even orders, and were generated not only by linearly but also by circularly polarized laser pulses. These observations excluded atomic harmonics via the recollision process [54] from possible mechanisms. Further, the combs with similar properties and brightness were observed in a relatively wide range of plasma densities, with the maximum-to-minimum ratios of ~3 to 5, which also excluded the precise phase matching required for efficient atomic harmonic generation [55].

Electrons with broad spectra were accelerated simultaneously up to a few hundred MeV; however, the observed combs were not correlated with the electron properties. Further, the fringe periodicity was in the optical region, ~1 eV. Together with the temporal coherence, these observations excluded incoherent betatron radiation [56] determined by the plasma frequency and electron energy.

The high brightness of the new x-ray source allowed to record the spectra in a single laser shot. Even conservatively-estimated absolute photon yield exceeded by orders of magnitude nonlinear Thomson scattering [10] estimated under the most favorable conditions.

Finally, the x-ray yield scaled quadratically with the driving laser power, Fig. 3.

Thus, the experimental observations could not be explained by the mechanisms known before: a new mechanism was necessary to explain the discovered x-ray source.

Experimental hints to the new mechanism included:

(i) observation of both odd and even orders, Fig. 2, which indicated temporal coherence and suggested an absence of the electric field reversal symmetry [57] in the x-ray source, and therefore possibly off-axis emission source;
(ii) red-shifted base harmonic frequency, Fig. 2, consistent with gradual laser frequency downshift [58, 59] as the laser pulse, during its nonlinear propagation, spends its energy on the excitation of plasma waves (which is connected with the depletion length in laser electron accelerators [60, 61]): this suggested that the x-ray source was near the head or center of the laser pulse, as the tail of the laser pulse is blue-shifted due to photon acceleration [62-64];
(iii) observation of bright emission, Fig. 2, with the x-ray yield much higher than expected from nonlinear Thomson scattering that can be accumulated from all the electrons encountered by the laser pulse, which suggested that the emission scaled faster-than-linearly with the electron number, i.e. there was constructive interference (coherent emission) in the x-ray source;
(iv) observation of deep (up to 100% visibility) modulations, Fig. 2(b) indicating the presence, with a linearly polarized laser, of two equal mutually coherent sources separated in space and/or time;
(v) observation of spatially-coherent EUV radiation (photon energy >14 eV) at up to 8° off-axis angle; the coherence properties were consistent with a point-like source model [47].

## 3. Theoretical explanation and model

The experimental results presented above are explained by the BISER mechanism.

**PIC simulations.** Particle-in-cell (PIC) simulations reveal a typical scenario of the relativistically strong ($a_0 \gg 1$) laser pulse interaction with underdense ($n_e/n_{cr} \ll 1$) plasma (see [48] and supplementary therein). The laser pulse undergoes relativistic self-focusing [65-67] accompanied by the amplitude increase and spot narrowing (this is the simplest case of no-filamentation; for that, the laser pulse waist at focus in vacuum should be less than the characteristic wavelength of plasma [37], $D < \lambda_{pe} = 2\pi c/\omega_{pe}$). The self-focused laser pulse pushes the plasma electrons out, thus forming a nearly electron-free *cavity* [68-70]. Due to the finite transverse dimension of the self-focused laser pulse the electrons are pushed transversely as well. When the laser pulse and thus the *cavity* become sufficiently narrow, $D < 4ca_0^{1/2}/\omega_{pe}$, the transversely pushed electrons form an outgoing *bow wave* [37]. The walls (outlines) of the *cavity* and *bow wave* have substantially higher electron density than the initially unperturbed plasma, Fig. 4. Much higher electron density is in the *electron density spike* at the *joint* of the walls of the *cavity* and *bow wave*. The *bow wave* detaches from the *cavity* inside the laser electric field, thus the *electron density spike* is driven by the laser field, therefore it emits high-order harmonics. In more detail, electrons flowing through the *density spike* and undergoing acceleration emit a broadband spectrum up to high frequencies; this emission, driven by the laser, is (almost)periodic [71, 72], thus, the spectrum is modulated with optical frequency.

The moment of harmonic generation is shown in Fig. 4, that presents the results of the two-dimensional (2D) PIC simulation of the idealization of an already self-focused Astra Gemini laser pulse (linearly-polarized along the $y$-axis) with $a_{0\,\max} = 10$, duration of $\tau = 16\,\lambda/c \approx 43$ fs, and full width at half maximum (FWHM) waist of $D = 10\lambda = 8$ μm, interacting with a plasma of electron density $n_e = 0.01 n_{cr} = 1.73 \times 10^{19}$ cm$^{-3}$. The simulation box was $87\lambda \times 72\lambda$, with the resolution of $\lambda/1024$ along the $x$ (propagation) axis and $\lambda/112$ along the $y$ axis; the PIC-defined quasiparticle number was $6 \times 10^8$.

Here and below ions were assumed immobile due to their inertia; in other terms, the laser amplitude is well below the threshold for a substantial contribution from the ion dynamics, $a_0 \ll \sqrt{m_i/m_e}$ ($m_i$ is the ion mass), and the timing of the high-frequency radiation generation is negligible in comparison with the ion response. The simulations were performed with the code REMP [73] using periodical transverse and absorbing longitudinal boundaries.

As seen in Fig. 4(a), the laser pulse creates the electron-free *cavity* and *bow wave* in the originally uniform plasma. The dense and sharp walls of the *cavity* and *bow wave* are modulated by the laser pulse field; the modulations actually portray the phase of the laser pulse field; in particular, the modulation wavelength is the same as the local wavelength of the evolving laser pulse.

Fig. 4(b) shows the electron density lineout across the *electron density spike* at the *joint* of the walls of the *cavity* and *bow wave*; this data is taken 10 laser cycles earlier than in (a) in order to demonstrate the structure of the spike responsible for the emission of the high-frequency radiation within the spectral range from $60\omega$ to $100\omega$ shown in (a) in terms of the electromagnetic energy density $W$ (the red colorscale). We identify this high-frequency radiation as BISER; it appears as a train of ultrashort pulses (see Section **7**) propagating about the direction of the driving laser pulse.

The profile of the *electron density spike* in a log-log scale is shown in Fig. 4(c). It is remarkable that the density grows as $n_e(y) \propto (y - y_C)^{-2/3}$ when the transverse coordinate approaches the spike location, $y_C$, down to the resolution limit of $|y - y_C| < 0.02\lambda$, which is determined by the size of the PIC-defined quasiparticle form-factor. This power-law density dependence on the coordinate at the joining of two density walls, caused by a multi-stream plasma flow, is very characteristic and gives an important clue to the mechanism of the *electron density spike* formation, as explained below with catastrophe theory.

The high-frequency radiation shown in Fig. 4(a) has a harmonic spectrum extending up to the ~128$^{\text{th}}$ harmonic order [48], as is allowed by the simulation resolution. The emission is slightly off-axis, with the emission cone size of several degrees. The greater the harmonic order, the less the corresponding off-axis angle is. The harmonic energy yield obtained in the simulation is close to the experimental data obtained with the Astra Gemini laser.

A three-dimensional (3D) structure of the walls of the *cavity* and *bow wave* and the source of harmonic generation are shown in Fig. 5(a) which presents the results of the 3D PIC simulation for another case of the idealization of an already self-focused Gaussian laser pulse (linearly-polarized along the $y$-axis) with $a_{0\,\max} = 6.6$, duration of $\tau = 10\,\lambda/c$, and full width at half maximum (FWHM) waist of $D = 10\lambda$, interacting with a plasma of electron density $n_e = 0.001 n_{cr}$. The simulation box was $125\lambda \times 124\lambda \times 124\lambda$ with periodical transverse and absorbing longitudinal boundaries, with the resolution of $\lambda/32$ along the $x$ (propagation) axis and $\lambda/8$ along the $y$ and $z$ axes; the PIC-defined quasiparticle number was $2.3 \times 10^{10}$.

BISER shown in Fig. 5(a), in terms of the electromagnetic energy density $W$ (red crescents), is emitted from two opposite locations arranged along the laser field polarization ($y$ axis). Indeed, the electric field pushes electrons along the polarization direction, therefore a higher electron density is built up along that direction. Further, in the case of constructive interference, the emission intensity quadratically depends on the number of emitters, therefore regions with higher electron density are better seen (because they are more intense). As a result, we observe a double source of BISER. We note that Fig. 5(a) contains only few harmonics due to the limited resolution of the 3D simulation; a more detailed analysis shows that the higher the emitted harmonic order the smaller is the emission source; the same effect is much better seen in high-resolution 2D simulations. Naturally, lower orders are emitted from larger sources; in particular, the second harmonic [74] is emitted not only from the cusps but also from the cavity walls [75, 76].

The electron density distribution in Fig. 5(a) demonstrates the result of a multi-stream motion of the electron fluid: the laser pulse expels electrons in both the longitudinal and transverse directions while the electrostatic potential of the non-compensated background of slowly responding ions (in simulations, ions actually were assumed immobile) pulls the electrons back. Fig. 5(b) shows how electrons originally located in the plane $\Pi_0 = (x, y, z = 0)$ get shifted along the $y$ axis. The PIC-defined quasiparticles, representing electrons, are well-ordered according to their initial position $y_0$ (the order is represented by colour). They are drawn in the 3D space $(x, y, y - y_0)$. Here $y - y_0 = \int (p_y/m_e\gamma_e)dt = \langle p_y/\gamma_e\rangle/m_e$ is the displacement along the $y$ axis from the initial coordinate $y_0$; $p_y$ and $\gamma_e$ are the electron transverse momentum and Lorentz factor, respectively. We see that the initially flat surface, consisting of electrons at rest, is stretched and modulated by the laser field so that the surface becomes folded. The resulting electron density in the plane $\Pi_0$ is approximated by a projection of the folded surface onto $\Pi_0$, as exemplified in Fig. 5(c) for a model surface. The folds exactly correspond to the walls of the *cavity* and *bow wave*; the joint of two folds exactly corresponds to the *electron density spike*. Obviously, particles at the most curved part of a fold are much closer to each other (in the real space) than in the almost flat parts. Also, the folds have a finite extent along the $y$ axis for a fixed value of $x$, therefore its projection onto $(x, y)$ plane has an abrupt end. The corresponding abrupt increase of the electron density at the walls of the *cavity* and *bow wave* is a catastrophe of the plasma flow.

**Catastrophe theory model.** The formation of a strongly localized *electron density spike* and its peculiar properties, namely robustness to oscillations imposed by the laser field and the $(-2/3)$ power-law density dependence, Fig. 4(c), can be explained with the help of catastrophe theory [38, 39].

The formation of catastrophes is illustrated by the following one-dimensional (1D) model of electron motion along the transverse coordinate $y$, Fig. 6(a). Initially, the electrons with spatially varied momentum $p_y$ homogeneously fill the $y$ axis. In the phase space $(y, p_y)$, the phase "volume" occupied by the electrons is represented by a parametric curve $S(t, y, p_y) = 0$, with a parameter $t$ (an analogue of time). Due to the momentum inhomogeneity, the order of electrons eventually changes, thus a multi-stream flow arises and the curve $S$ gets folded. The electron density as a function of $y$ is obtained by a projection of the electron phase space onto the $y$ axis,

$$n_e(t, y) = \int_{-\infty}^{+\infty} f(t, y, p_y)dp_y, \quad (1)$$

where $f(t, y, p_y) \propto \delta[S(t, y, p_y)]$ is the particle density in the phase space $(t, y, p_y)$ and $\delta$ is the Dirac delta function. At the position of a fold, the density rises sharply. The emergence of a singularity in a continuous flow is called a catastrophe. Note that in the case of the fluid approximation the projection (1) gives an infinite density, while in the case of a finite particle number the density, although exhibiting a sharp rise, remains finite.

The singularity type is determined by the structure of the phase "volume". In general case, a fold can be approximated by a portion of a parabola,

$$y_F - y \propto (p_{yF} - p_y)^2, \quad (2)$$

Fig. 6(c), while the portion near the curve $S$ inflection can be approximated by a cubic,

$$y_C - y \propto (p_{yC} - p_y)^3, \quad (3)$$

Fig. 6(b). Therefore, the projection (1) gives the density singularity of

$$n_{\mathrm{eF}} \propto \frac{\partial S/\partial y}{\partial S/\partial p_y} = \frac{dp_y}{dy} \propto (y_F - y)^{-1/2} \quad (4)$$

at the fold and of

$$n_{\mathrm{eC}} \propto (y_C - y)^{-2/3} \quad (5)$$

at the inflection point. The approximations of an analytical curve at a fold (extremum) or at inflection, respectively, by a parabola or a cubic are the simplest ones that are robust to smooth perturbations.

In a higher dimensionality space $(t, y, p_y)$, the inflection point corresponds to a merger of two folds of the surface, Fig. 6(d). Near the merger point, $(t_C, y_C, p_{yC})$, the surface $S$ can be approximated implicitly by

$$(p_y - p_{yC})^3 + \alpha(t - t_C)(p_y - p_{yC}) + \beta(y - y_C) = 0$$

with some constants $\alpha \neq 0$ and $\beta \neq 0$ ([39], Chapter 5 paragraph 2), in analogy with the approximations by a parabola and a cubic used above. Here the projection onto the $(t, y)$ plane produces two singular lines joining at a point with a singularity of higher order. If we apply small smooth perturbations to the electron phase "volume" represented by the curve $S$, the fold and the inflection still exist and can be approximated locally by a parabola and a cubic, respectively. In other words, the corresponding singularities are robust with respect to perturbations. The rigorous proof of this fact is given in catastrophe theory ([38], Chapter 2; [39], Chapter 6, paragraphs 4 and 5), which is a well-established branch of mathematics studying singularities formation in dynamical systems and in mappings. Here catastrophe theory establishes two *structurally stable singularities*, the *fold catastrophe* ($A_2$, according to V. I. Arnold's classification) and the *cusp catastrophe* ($A_3$).

Catastrophes of the same types as in the model in Fig. 6(d) appear in the laser plasma during the cavity and bow wave formation, Fig. 5(a,b). Here the $x$ coordinate of the laser pulse propagation corresponds to the reversed time in Fig. 6. The cusp catastrophe appearing at the joining of the *cavity* and *bow wave* walls is the *electron density spike* seen in the 3D PIC simulations, Fig. 5(a). The *spike* is located in a ring surrounding the cavity head, moving together with the laser pulse. Being pushed by the electromagnetic field of the laser pulse the electrons form a flow modulated with the local laser frequency. Correspondingly, the *spike* undergoes periodic oscillations and emits electromagnetic radiation with high-order harmonics of the local laser frequency. Catastrophe theory ensures that the described singularities are structurally stable which means that they survive perturbations imposed by the laser pulse and other factors such as electron density fluctuations. Singularities sharper (by orders of magnitude denser) than the cusp can momentarily appear, however, they are not structurally stable against the perturbations (the lifetime of such singularities is likely much shorter than the wave formation time, therefore their contribution to radiation is likely small – yet, this needs to be tested experimentally). Other types of structurally stable singularities in collisionless plasma are discussed in Refs. [36, 77]. Structurally-stable relativistic plasma singularities reflecting light with frequency upshifting, duration shortening, and intensification due to the double Doppler effect are known as Relativistic Flying Mirrors in various regimes [18, 19, 78-87].

We note that the sharpness of the singularities plays a crucial role in generating the highest-frequency radiation, consistently with the conjecture discussed in the Introduction. The initial plasma temperature, or other reasons causing widening of the initial electron phase space, will result in reduced sharpness [88, 89]. Comparison of electron density distributions in PIC simulations with initially cold and hot (30 keV) plasmas is shown is Fig. 1(b,c) of Ref. [90].

## 4. Model validation in experiments

As seen in the previous Section, the catastrophe theory model explained the existing experimental observations. In addition, the model made three testable predictions:

(i) BISER originates from point-like (nanometer-scale) x-ray sources,

(ii) For a linearly polarized laser, there exist two main point-like BISER emitters located along the laser polarization,

(iii) The off-axis emission is BISER's main component: it is not limited to the low-photon-energy (~14-30 eV) spectral region, but extends up to hundreds of eV, and at these photon energies can cover the angle of up to 10-20°.

These predictions were tested in the subsequent experiment [23] using the J-KAREN laser, with the peak power up to 20 TW and vacuum irradiance of up to $10^{19}\ \mathrm{W/cm^2}$. The emission from plasma was collected by a normal-incidence spherical XUV mirror with its center located at 13° off-axis (in the polarization plane) and the half-cone acceptance angle of 5°. Thus, the mirror covered the 8 to 18° range in the polarization plane, and 0±5° in the perpendicular plane. The mirror had an aperiodic Mi/Si multilayer coating [91-94] reflecting radiation with photon energy up to 100 eV (wavelength 12.4 nm, L edge of absorption of Si). Optical blocking filters were used to stop the laser and low-frequency plasma radiation, so that the observable photon energies were above ~60 eV (wavelength shorter than ~20 nm). The mirror imaged the BISER source on one of the two detectors, with magnification from ~5 to 7.

The first detector was a back-illuminated x-ray Charge Coupled Device (CCD) equipped with a transmission grating, thus constituting an imaging (spatially-resolving) XUV spectrograph (similar to employed in [95, 96]). It revealed the double BISER source aligned along the laser polarization direction, Fig. 7 and Fig. 8(a,b), i.e. it confirmed the prediction (ii). The spectra extended up to the spectrograph throughput cutoff at 100 eV ($L_{Si}$ edge), Fig. 8(c), thus confirming the prediction (iii). However, with the XUV image magnification of 6.85 and CCD pixel size of 13.5 μm, the prediction (i), of nanometer-scale sources, could not be tested.

For the second detector a high-resolution LiF crystal [97-99], was used. It consisted of an imager with a magnification of 5.53 (the XUV mirror was mounted on an in-vacuum motorized XYZ translation stage which controlled the image position on the CCD or LiF surfaces). An example of a single-shot experimental image is shown in Fig. 8(d). Like the CCD detector, LiF images showed the double point source aligned along the laser polarization. It also showed that the source must be spatially coherent, otherwise the fringes visible in Fig. 8(d) would be blurred out. We emphasize that this spatial coherence could not be explained by the propagation effect (van Cittert-Zernike theorem [100]), but rather was a direct consequence of the source coherence [23].

Most importantly, the high resolution of the LiF detector allowed a better estimate of the source size. The directly shown (apparent) size was ~0.8 μm, which was limited by the aberrations of the spherical mirror operating at somewhat non-normal incidence. Physical optics propagation modelling with different source sizes showed that the source size was not larger than 100 nm. Thus, the prediction (i) was also confirmed in the experiment, which validated the catastrophe-theory-based model of BISER.

## 5. Proposed observation of relativistic plasma singularities

As we see, a relativistically strong tightly focused laser pulse easily creates electron density singularities in underdense plasma; these singularities move as fast as the laser pulse, and even faster; some of them dwell inside the laser field or remember modulations imposed by the laser field on the plasma flows that create the singularities. The existence and motion of the electron density singularities can be observed experimentally by diffraction of an auxiliary ultrashort probe laser pulse on the singularities [101, 102]; such the diffraction is also a manifestation of the BISER effect. The diffracted radiation can be extracted by Schlieren imaging [103, 104], described in the following scheme, Fig. 9(a).

**Schlieren imaging.** The driver laser pulse creates electron density singularities; at this moment the probe laser pulse irradiates the scene in the transverse direction. Obviously, the probe pulse should be weak (not affecting the plasma dynamics) and sufficiently short (to prevent or decrease blurring due to the singularity motion). The probe pulse is diffracted off the singularities. The geometric optics approximation is not applicable because the singularities are strongly localized. The probe pulse together with the diffracted radiation goes into a special optical system which makes a Schlieren image, Fig. 9(a). A point-like singularity produces a spherical diffracted wave, which is imaged by the lens as a point-like object, while an opaque plate blocks the non-refracted and slightly-refracted radiation from smooth density distortions; in this way the singularities can be identified.

The effect of the cusp singularity on the probe pulse phase can be roughly estimated assuming a stationary cusp density distribution, Fig. 9(b):

$$n_e(y) = n_{e0}(\lambda/y)^{2/3}. \quad (6)$$

Here $\lambda$ is the probe wavelength, and $y$ is the coordinate along the probe pulse propagation direction. The phase shift is

$$\Delta\varphi = (2\pi/\lambda)\int_0^{M\lambda}(\mathfrak{n}-1)dy, \quad (7)$$

where $\mathfrak{n}$ is the refractive index:

$$\mathfrak{n} = \sqrt{1-n_e/n_{cr}}. \quad (8)$$

The integration from $y_{\min} = 0$ to $y_{\max} = M\lambda$, where $M$ is the dimensionless length, gives the real part

$$\Re[\Delta\varphi] = 2\pi\left(\left(M^{2/3} - n_{e0}/n_{cr}\right)^{3/2} - M\right), \quad (9)$$

which slowly diverges as $\Re[\Delta\varphi] \rightarrow -3\pi M^{1/3}\, n_{e0}/n_{cr}$ at large $M$. In 3D, the electron density decreases quickly for $y > D$ (because gas is not yet ionized there), so for estimation one can assume $M = D/\lambda$. The imaginary part is

$$\Im[\Delta\varphi] = 2\pi(n_{e0}/n_{cr})^{3/2}, \quad (10)$$

corresponding to the transmission intensity of $\exp\left(-4\pi(n_{e0}/n_{cr})^{3/2}\right)$. For the cusp parameters as in Ref. [23], $n_{e0} = 0.023n_{cr}$, this gives the real part of $\Re[\Delta\varphi] = -0.46$ rad, and the imaginary part of $\Im[\Delta\varphi] = 0.02$ rad, corresponding to the beam transmission of 0.96, i.e. 4% attenuation. Such a phase shift and attenuation are well within measurable ranges.

**PIC simulations.** In the case of a linearly polarized laser pulse, soft x-rays are emitted by a pair of point-like sources, as predicted in Section 3 and demonstrated in Section 4. The corresponding pair of density singularities (cusps) are seen in 2D PIC simulations illustrating the possibility of Schlieren imaging. 2D PIC simulations demonstrate two important properties of the diffracted probe, namely, the well discernible spherical wavefronts from the electron density singularities and the expected Doppler shift of the diffracted radiation frequency due to the singularities fast motion. Fig. 10 shows a propagation of the probe pulse through plasma where the driver pulse creates electron density singularities. Both the pulses have the same wavelength $\lambda_0$. The initial electron density is $n_{e0} = 0.01n_{cr} \approx 1.1\times10^{19}\text{cm}^{-3}\times(1\ \mu\text{m}/\lambda_0)^2$.

The Gaussian $p$-polarized driver pulse propagating along $x$ axis has the dimensionless amplitude of $a_0 = 6.6$, focal spot size of $5\lambda_0$, FWHM length of $5\lambda_0$, irradiance of $6\times10^{19}$ W/cm$^2\times(\lambda_0/1\ \mu\text{m})^2$, power of 15 TW, and pulse energy of 240 mJ $\times(\lambda_0/1\ \mu\text{m})$.

The Gaussian $s$-polarized probe pulse propagating along $y$ axis has the dimensionless amplitude of $a_{0p} = 0.01$, focal spot size of $(160/3)\lambda_0$, FWHM length of $3\lambda_0$ (we note that a probe pulse duration of 6 fs, which corresponds to ~2 laser cycles, is already available in experiments [105]).

The $p$-polarized driver and $s$-polarized probe have the electric field vector directed, respectively, along $y$ and $z$ (perpendicular to the simulation box plane) axis. This difference helps to filter out plasma emission due to the driver pulse. In the plane perpendicular to the driver pulse polarization, emission of electrons dragged by the driver is much less efficient, therefore it makes much less impact on the imaging by the probe.

The simulations were performed with the code REMP [73] using two stages: with and without the driver pulse; subtracting the electromagnetic field s-polarized component obtained in the latter case from that obtained in the former case gives the pure diffracted radiation. Additionally, the blocking technique in the Schlieren imaging was simulated by applying a Fourier filter, which removes modes propagating within $\pm9°$ about the direction of the probe pulse.

In Fig. 10, the driver pulse can be traced by the electron density modulations with the laser wavelength at the front of the leading wake wave cavity. The laser pulse is intense enough to accelerate electrons so that they cannot be bounded in the Langmuir oscillations. Therefore, the wake waves break. The resulting multi-stream flows of electrons create various density singularities seen as the low-dimensionality regions of high electron density. In particular, (one-dimensional) thin high-density shells (e.g., the walls of the *cavity* and *bow wave*) correspond to the fold singularity. The (zero-dimensional) points joining these thin shells correspond to the cusp singularity. We note that in a three-dimensional configuration, the mentioned examples of fold and cusp singularities are two- and one-dimensional, respectively, as in Section 3; and other types of structurally stable singularities appear, e.g. the swallow tail singularity [36].

A prominent diffraction occurs on the cusps at the head of the leading cavity at $t = 105$ and $t = 120$. The diffracted light from a single cusp appears in the form of short pulses with characteristic spherical wavefronts slightly shifted with respect to each other along the $x$ axis due to the cusp motion (a manifestation of the Doppler effect). Correspondingly, the diffracted light from a single cusp is blue shifted on the right and red shifted on the left. Spherical fronts revealed at $t = 150$ correspond to various density singularities encountered by the probe pulse.

In addition to perfectly spherical fronts, there are initially flat finite fronts of the probe reflected from the walls of the *cavity* and *bow wave*. They look like comet-like patches on the spherical fronts from cusps. This type of reflection is the same as in Refs. [19, 106], where the probe pulse (there called *source pulse*) is counter-propagating with respect to the driver pulse; due to modulations, the *cavity* and *bow wave* walls act as a thin *diffraction grating* flying with relativistic velocity.

The characteristic size of the dense regions corresponding to the fold and cusp singularities is much smaller than the probe wavelength. Thus, the electromagnetic emission from electrons in these regions, induced by the probe pulse, is temporally and spatially coherent for wavelengths about $\lambda_0$ (actually, even for much shorter wavelengths, of the order of the least characteristic size of the high-density region). Further, these regions consist of electrons evacuated from the cavity. For the case of $\lambda_0 = 1$ μm, we can roughly estimate the number of electrons from the cavity as $N_e = (4\pi/3)(10\lambda_0)^3 n_{e0} \approx 4.6 \times 10^{10}$ . Considering the quadratic dependence of the coherent emission intensity on the number of emitters, it is not surprising that the probe pulse undergoes efficient diffraction on the density singularities.

The diffracted light has interesting angular spectra where one can distinguish the diffraction on the steady and moving parts of plasma [79, 80], including the characteristic Doppler shift.

Even a not-so-ideal probe pulse containing satellite pre- and post-pulses can be useful for Schlieren imaging of fast density singularities, as shown in Ref. [102]. This is possible if the probe pulse has high enough contrast between the main pulse and the satellites and background.

**Lampa–Penrose–Terrell effect.** The image of the constellation of singularities will be such as if the singularities were rotated with respect to each other, Fig. 10, Fig. 11. This phenomenon is similar to the Lampa–Penrose–Terrell effect [107]. The singularities move with the velocity approximately equal to the group velocity of the driver pulse, which is close to the speed of light in vacuum, $c$. Thus, the constellation of singularities significantly drifts during the probe pulse propagation through it. Therefore, parts of the constellation, which are irradiated later, appear in the image at positions more shifted in the direction of the constellation drift. This is seen in Fig. 10.

If the cusps were at rest, the two spherical waves from them would have their centers exactly aligned along the probe pulse propagation, thus we would see just a single image of the diffracting object, Fig. 11(a). However, due to the fast motion of the cusps, the probe pulse meets them at different horizontal position. Overall, this results in the apparent rotation of the pair of cusps as a whole, so that we will see two separated diffracting objects, Fig. 11(b). For this, the probe pulse must be short enough (or the main-to-satellite pulse contrast must be significant), as stated above. In our case the probe pulse is shorter than the constellation size, which is about the waist of the driver pulse along the $y$ axis. Fig. 11(c) sketches out how the cusp is imaged in a 3D geometry. The cusp singularity appears in the form of a ring, Fig. 5(a). Due to the linear polarization of the driver, more electrons concentrate in the direction of the driver polarization. Since regions with higher electron density diffract light more efficiently, a sort of ellipse with a higher signal magnitude at two opposite spots appear.

## 6. Driving laser requirements and parameters

Several experimental campaigns performed with similar laser parameters and similar or even identical detectors (spectrographs) showed significantly different BISER x-ray yields, Fig. 12. Such orders-of-magnitude-difference was surprising, considering that in each campaign the BISER generation was optimized by varying the plasma density and gas jet position. Subsequent analysis showed that the apparently random data produced well-established trends with respect to the laser quality, which showed that the observed randomness was due to laser fluctuations. In particular, the focal spot quality significantly affected the results [108, 109], strongly favoring higher-quality spots. The physical process behind this is laser pulse filamentation, as suggested by observation of the output (after the plasma) laser beam profile and angular splitting of the BISER radiation [108].

Among the factors influencing the focal spot, the two most important ones are the wavefront distortions and spatiotemporal couplings in the laser system.

The wavefront distortion can be characterized by σ, the root-mean-square (RMS) deviation from a perfect spherical wavefront. The laser pulse irradiance can be estimated as $I = I_0 \exp[-(\eta\pi\sigma/\lambda)^2]$, where $I_0$ is the ideal irradiance achieved at $\sigma = 0$ (the diffraction-limited irradiance), $\lambda$ is the laser wavelength, and the factor $\eta \approx 2$ [110]. For efficient BISER generation, $\sigma < \lambda/8$ is at least required [108], which practically coincides with the Maréchal criterion of a diffraction-limited beam.

Spatiotemporal couplings can drastically degrade the laser focusability even in the case of a perfect wavefront [111-113]. In particular, the angular dispersion, i.e. dependence of the propagation direction on wavelength, is a common problem of CPA lasers arising from a misalignment of the stretcher and/or compressor, or from propagating the laser beam through wedged optics and prisms. Angular dispersion causes an increase of both the focal spot size and pulse duration [111]. To keep these effects at a minimum, the divergence due to the angular dispersion should be much smaller than the diffraction divergence: $C\Delta\lambda \ll$

$\lambda/D_0$, where $C$ [μrad/nm] is the linear angular chirp (assumed to be the dominant angular dispersion term), $\Delta\lambda$ is the laser bandwidth, and $D_0$ is the laser beam diameter. Thus, this requirement is especially stringent for high-power femtosecond lasers, which have simultaneously wide bandwidth and large beam diameter. For example, for the J-KAREN-P laser ($D_0 \approx 280$ mm, $\Delta\lambda \approx 40$ nm, $\lambda \approx 0.8$ μm) this gives $C \ll 0.075$ μrad/nm; for 90% of the ideal irradiance, $C < 10^{-2}$ μrad/nm is required. This translates into the compressor grating angle alignment accuracy requirement of ~10 μrad [108], for all three angles of each of the compressor gratings. It is this accuracy that is achieved in the J-KAREN-P compressor [114].

Other spatiotemporal distortions, such as chromatic aberration and propagation time difference [115, 116] caused by non-achromatic lenses, should be eliminated as well. Examples of these aberrations caused by singlet-lens expanders, and their elimination by mirror-based expanders, are provided in Ref. [114].

Finally, we note that high-power lasers are often characterized by an over-idealized "Strehl ratio" $S_{WF}$ determined from the measured wavefront, FWHM spot diameter, and FWHM pulse width. While these parameters can sometimes be useful, they are usually misleading and do not give an adequate understanding of the laser performance in experiments.

The over-idealized "Strehl ratio" $S_{WF}$ determined from the measured wavefront, i.e. the peak value of the calculated point-spread function normalized to its value for wavefront-distortion-free beam, does not include spatiotemporal couplings. Further, even state-of-the-art wavefront sensors cannot measure high-frequency wavefront distortions. Nor do they measure the wavefront at the low-intensity edges of the beam, which contribute significantly to the spot size and quality: note that the focal spot in the Fraunhofer diffraction case, which is a good approximation for nearly all experiments, is determined by the Fourier transform of the near field, i.e. square root of the intensity, and the points further away from the axis (in the near field) contribute more strongly to the on-axis intensity (in the far field). Thus, typical values determined from the wavefront measurement, $S_{WF}$~0.9 or even higher, usually have no relevance to the actual laser performance.

Likewise, FWHM values of the focal spot diameter $D_{\rm FWHM}$ and pulse width $\tau_{\rm FWHM}$ are often used to estimate the laser irradiance as $I = 4P/\pi D_{\rm FWHM}^2$ and power as $P = E_L/\tau_{\rm FWHM}$. However, these estimates do not take into account low-intensity parts of the irradiance or power distribution with the intensity lower than 50% of the peak, while a significant portion of the total laser energy can be spread into the low-intensity halo around the spot, or low-power satellite pulses before and/or after the main pulse. In some cases, rather typical for CPA lasers, the peak irradiance can be overestimated by an order of magnitude or more. Examples are provided in Fig. 1(a) and Section 8 of Ref. [114]. In addition, processes like filamentation can degrade the performance of low-quality lasers even further, as mentioned above.

The reliable and relevant characteristics of the laser quality are the effective pulse width (duration) and the effective spot area (or the effective spot radius) [114]. The effective pulse width (duration) is defined as

$$\tau_{\rm eff} = \int p(t)\mathrm{d}t,\ p(t) = \frac{P(t)}{P_0}, \tag{11}$$

where $P(t)$ is the power dependence on time, $p(t)$ is the normalized dependence, and $P_0$ is the peak power. The effective pulse width (duration) equals the area under the normalized power curve $p(t)$.

Similarly, the effective spot area and radius are defined as

$$A_{\rm eff} = \pi r_{\rm eff}^2 = \iint f(x,y)\mathrm{d}x\mathrm{d}y,\ f(x,y) = \frac{F(x,y)}{F_0}, \tag{12}$$

where $F(x,y)$ is the fluence distribution, $f(x,y)$ is its normalized distribution, and $F_0$ is the peak fluence. The effective spot area equals the volume under the normalized fluence surface $f(x,y)$.

The peak power and irradiance are

$$P_0 = \frac{E_{\rm L}}{\tau_{\rm eff}},\ I_0 = \frac{P_0}{A_{\rm eff}} = \frac{P_0}{\pi r_{\rm eff}^2}. \tag{13}$$

The ideal (diffraction-limited) focal spot can be calculated as described above, i.e. as a Fourier transform of the square root of the intensity (fluence) distribution measured before the final focusing optics (so-called near field). The ratio of the effective areas of the ideal and real focal spots gives a meaningful definition of the Strehl ratio:

$$S = \frac{A_{\rm eff,DL}}{A_{\rm eff}} = \frac{r_{\rm eff,DL}^2}{r_{\rm eff}^2}. \tag{14}$$

The Strehl ratio determined in this way is usually notably smaller than the one determined from the measured wavefront [108, 114], however, it is much more relevant and indicative of the performance of the laser in real experiments [109].

Concluding this Section, we note that experiments with higher laser quality produced much brighter and much more stable BISER. We envision further BISER enhancement with future quality and stability improvements of high-power lasers.

## 7. On the BISER pulse duration

The temporal shape of BISER is yet to be measured. However, several conclusions can be deduced from the existing experimental data and simulations.

First, the broad bandwidth of the BISER spectra suggest that the minimum achievable pulse durations can be very short. The fundamental Uncertainty Principle, $\tau\Delta\omega \geq 2\pi$, states that the minimum pulse duration can be estimated as $\tau_{\min} = \frac{2\pi}{\Delta\omega} = \frac{\omega}{\Delta\omega}\frac{2\pi}{\omega} = N_{\min}T_0$, where $T_0 = \frac{2\pi}{\omega}$ is the x-ray wave period (one cycle) and $N_{\min} = \omega/\Delta\omega$ is the minimum number of cycles. More precisely, the minimum pulse duration can be found by the Fourier transformation of the spectral amplitude (i.e. the square root of the spectral intensity) assuming zero spectral phase, i.e. perfect phase matching. This standard procedure gives the so-called *transform-limited (TL) pulse*. As the BISER spectra are essentially "white", i.e. $\Delta\omega/\omega \sim 0.5$-1, BISER can potentially generate *ultimately-short single-cycle x-ray pulses*.

Second, the BISER spectra exhibit complex spectral shapes, Fig. 2 and Fig. 8(c). Large-scale modulations, with the period from several units to several tens of eV, indicate the presence of temporally coherent pulses with attosecond separation on the detector [117], while small-scale modulations with the period from ~eV to sub-eV indicate temporal coherence on the femtosecond time scale [71, 72]. These spectral features indicate a complex overall temporal shape.

Third, similarly to other harmonic generation mechanisms [118-120], PIC simulations [23] demonstrate that BISER from moving oscillating electron density spikes has a form of attosecond pulse trains, Fig. 4(a) and Fig. 13(a), with the pulse duration indeed approaching the transform limit, dashed line in Fig. 13(b), thus confirming the two conclusions above. Fig. 13(b) also shows two transform-limited pulses calculated from the experimental spectra:

(1) the black line (150 as pulse, FWHM) calculated from experiment [23] with a 8th order super-Gaussian spectral filter centered at 76 eV photon energy with ±13 eV bandwidth, which is somewhat narrower than the bandwidth determined by the aperiodic multilayer Mo/Si mirrors [93] and Zr/Si multilayer filters [52] used in that experiment (60-100 eV);

(2) the red line (18 as FWHM, 22 as effective width) calculated from experiment [108] with a 6th order super-Gaussian filter centered at 250 eV photon energy with ±100 eV bandwidth.

We note that the latter pulse is shorter than the atomic unit of time, $\tau_a = \frac{\hbar}{\alpha^2 m_e c^2} = 24$ attoseconds, where $\hbar$ is the reduced Plank constant, $\alpha \approx 1/137$ is the fine-structure constant.

Fourth, the BISER spectral shapes in the experiments fluctuated shot-to-shot, suggesting temporal shape fluctuations as well. Thus, at the present state of the laser stability and interaction control, the temporal pulse shape measurements should be single-shot-based. This requires significant energy per pulse: for example, single-shot attosecond pulse measurement was demonstrated with several tens to hundreds of μJ with soft x-ray FEL [121].

Present-day attosecond pulse measurement methods, either multi-shot (scanning) [119] or single-shot based [122], require a secondary gas jet target, where the attosecond pulse interacts with either its own copy or with a synchronized infrared pulse. Particles (electrons or ions) resulting from such interaction are analyzed to derive information about the pulse shape and duration. To test if the BISER attosecond pulses can be characterized in similar setups, we performed preliminary experiments where single-shot BISER was focused into secondary Xe gas jet; the resulting photo- and Auger electrons were collected and characterized by high-resolution time-of-flight Magnetic Bottle Electron Spectrometer (MBES) [123]. Fig. 14 shows the comparison of single-shot data with a multi-shot average: despite nearly identical overall energy spectra, the single-shot data exhibits fine structure with sub-eV scale, while it is absent in the 10-shot average. This comparison shows that the BISER pulses must indeed be measured in the single-shot mode; it also suggests that the experimentally achieved signal-to-noise level is sufficient for such measurements.

Small source size, short pulse duration, and high photon number per pulse make BISER a very bright radiation source. Comparison of the conservatively-estimated brightness of BISER in previous experiments with other XUV and x-ray sources is shown in Fig. 15. The BISER brightness exceeds synchrotrons by several orders of magnitude and is comparable with XUV FELs.

## 8. BISER wavelength scalability

The presented results of theory and simulations are formulated in terms of dimensionless parameters. If, in two different physical settings, all such parameters corresponding to the laser and plasma properties are the same, then the predicted results should be identical (if applicability conditions hold in both settings, i.e. there is no need to consider additional physical effects). For example, for two laser systems with different wavelengths, if the dimensionless parameters of the laser ($a_0$, $\tau_{\mathrm{eff}} c/\lambda$, $r_{\mathrm{eff}}/\lambda$, F-number, focal spot shape, etc.) and plasma ($n_e/n_{cr}$, $m_i/m_e$, density profile, etc.) are similar, then the expected effects should be similar (e.g., relativistic self-focusing, an excitation of Langmuir and bow waves, wave breaking, a formation of density singularities, and BISER), both qualitatively and quantitatively. PIC simulations show the similarity of the plasma waves and singularities formation with the driving laser wavelength scaled to 4 – 40 μm [90, 124].

The presented experiments were performed with the laser wavelength of $\approx 0.8$ μm. A significant advantage in many respects may be gained with different laser wavelengths. For example, a longer laser wavelength may substantially ease measurements of density singularities and BISER pulse duration; a shorter laser wavelength can give access to more energetic photons in the BISER spectrum and to shorter BISER pulse duration (provided that the interaction regime is far from requiring inclusion of radiation reaction on individual electrons [125] and quantum effects [126]).

Recently a high-quality multi-terawatt $CO_2$ laser with the 9.2 μm wavelength was demonstrated at the BNL ATF facility [127, 128]. This long-wave infrared (LWIR) laser irradiating gas targets successfully produced the laser wake-field acceleration of electrons [129]. This achievement suggests that such the laser can also produce the electron density singularities and BISER.

An experimental probing of the electron density singularities is challenging, if both the driver and probe laser pulses have the wavelength of $\approx 0.8$ μm, because the singularities move with a relativistic velocity, have an apparent size of $\approx 0.1$ μm or smaller, and are separated by several micrometers (Section **5**). However, if the driver laser has a LWIR wavelength of 9.2 μm, the mentioned spatial parameters of singularities should be ~10 times greater, so it will be much easier to image them with the probe pulse of a Near-IR laser with the wavelength of ~0.8 or even 0.4 μm (its second harmonic), then a characteristic power-law of density near the singularities with the exponents of (-2/3) or (-1/2) may become measurable. We note that a synchronized sub-100 fs Near-IR laser is available at the BNL ATF facility [130]; it can be used as a probe for imaging the singularities.

Also, very challenging is the measurement of the BISER pulse duration (Section **7**). An alternative, although indirect, approach is to measure that duration when BISER is driven by a LWIR laser. Then, the BISER pulses will be in the infrared and visible spectral regions, so that they can be measured using one of the femtosecond pulse measurement techniques, e.g. [131-133]. In addition, other BISER radiation properties, which are challenging to measure in the XUV and x-ray spectral regions, can be measured in the wavelength-scaled experiments as well.

Table II compares the laser and plasma parameters describing the same model laser-plasma dynamics for two different laser wavelengths, $\lambda$; scalings show how the parameters depend on $\lambda$; the case of no dependence is encoded as $\lambda^0 = 1$.

**Table II. Parameters of lasers and plasma in the wavelength-scaled experiments**

| **Parameter** | **Scaling** | **NIR** | **LWIR** |
|---|---|---|---|
| Wavelength (μm) | $\lambda$ | 0.8 | 9.2 |
| **Dimensionless parameters:** | | | |
| Dimensionless amplitude (in vacuum) $a_{0,\mathrm{Vac}}$ | $\lambda^0$ | 1.4 | 1.4 |
| Effective pulse duration $\tau_{\mathrm{Eff}}/T_0$ | $\lambda^0$ | 16 | 16 |
| Focusing OAP *f*-number | $\lambda^0$ | *f*/9 | *f*/9 |
| Effective spot radius $r_{\mathrm{Eff}}/\lambda$ | $\lambda^0$ | 8 | 8 |
| Peak electron density $n_{\mathrm{e,max}}/n_{\mathrm{cr}}$ | $\lambda^0$ | 0.02 | 0.02 |
| Plasma wavelength $\lambda_{\mathrm{p}}/\lambda$ | $\lambda^0$ | 7.07 | 7.07 |
| Amplitude after self-focusing (estimated) $a_{\mathrm{SF}}$ | $\lambda^0$ | 3.7 | 3.7 |
| **Dimensional parameters:** | | | |
| Vacuum irradiance $I_{0,\mathrm{Vac}}$ (W/cm$^2$) | $\lambda^{-2}$ | $4.2\times10^{18}$ | $3.2\times10^{16}$ |
| Laser cycle duration $T_0$ (fs) | $\lambda$ | 2.7 | 31 |
| Effective pulse duration $\tau_{\mathrm{Eff}}$ (fs) | $\lambda$ | 43 | 491 |
| Effective spot radius $r_{\mathrm{Eff}}$ (μm) | $\lambda$ | 6.4 | 74 |
| Rayleigh length (μm) | $\lambda$ | 322 | 3700 |
| Peak power $P_0$ (TW) | $\lambda^0$ | 5.4 | 5.4 |
| Pulse energy $E_{\mathrm{L}}$ (J) | $\lambda$ | 0.23 | 2.6 |
| Critical density $n_{\mathrm{cr}}$ (cm$^{-3}$) | $\lambda^{-2}$ | $1.7\times10^{21}$ | $1.3\times10^{19}$ |
| Peak electron density $n_{\mathrm{e,max}}$ (cm$^{-3}$) | $\lambda^{-2}$ | $3.5\times10^{19}$ | $2.6\times10^{17}$ |
| Plasma wavelength $\lambda_{\mathrm{p}}$ (μm) | $\lambda$ | 5.66 | 65.1 |
| Irradiance after self-focusing (estimated) $I_{\mathrm{SF}}$ (W/cm$^2$) | $\lambda^{-2}$ | $2.9\times10^{19}$ | $2.2\times10^{17}$ |
| Nozzle diameter (mm) | $\lambda$ | 1 | 11.5 |

## 9. Future prospects

One of the most promising prospects of BISER is the realization of a bright coherent compact x-ray source that can generate powerful pulses with duration shorter than 24 attoseconds, i.e. shorter than the atomic unit of time. Why can such a source be interesting?

X-rays have a nanometer and shorter wavelength and thus, compared to optical imaging, provide much higher resolution, which is generally limited by wavelength (with an imaging-geometry-specific factor). In contrast to electron and atomic force microscopy, x-rays can image even thick and in-water samples, including live bio-samples "as they are" – without staining and other time-consuming and *structure-changing* steps. Photons are not charged and thus can be concentrated to intense ultrafast flashes providing high temporal resolution.

Recent x-ray imaging breakthroughs were triggered by two major developments: (i) bright x-ray sources, especially femtosecond XFELs and attosecond atomic harmonics [134], and (ii) lensless imaging methods, such as Coherent Diffractive Imaging (CDI) [135] and Holography [136], where the resolution is not limited by optics imperfections. Significant progress has been achieved, such as 7-nm resolution single-particle virus imaging [137] and 2-nm resolution metal nanoparticle imaging [138]. For the smallest samples, such as bio-molecules, the Image Before Destroy concept is assumed, where the x-ray pulse must be bright enough to provide single-shot exposure with sufficient signal-to-noise ratio [139], Fig. 16(a).

One next expected breakthrough is access to the atomic/quantum timescales determined by the atomic time unit, Fig. 16(b). Indeed, the quasi-classical “orbital period” of an electron in Bohr’s hydrogen atom is $2\pi\tau_a{\sim}150$ attoseconds; this duration can be considered as a time scale of complete loss of temporal resolution. Spatial resolution loss due to the "motion blur" during finite pulse duration should be determined from fully-quantum calculations, however, this method is yet to be developed; we can estimate it quasiclassically, Fig. 17. XFELs can achieve tens and hundreds of μJ energy per pulse (~$10^{11}$ to $10^{12}$ photons/pulse), but with a narrow bandwidth and the duration record of ~200-300 attoseconds [121] obtained by turning energy modulations into density spikes [140, 141]. (One can see that this method of density spike generation is based on the multistream flow of electrons in the beam, which is a typical case described by catastrophe theory. We therefore explain the attosecond XFEL pulses from these density spikes by the BISER effect.) The present record of an x-ray pulse duration of ~40 attoseconds is held by the atomic harmonics source [142]. However, the photon number per shot (or single-shot x-ray pulse energy) of such ultrafast sources is typically very low, e.g. in the record case, <$10^7$ photons/pulse at above-keV photon energies [4], which limits applications in science and technology. Thus, BISER has the potential to provide shorter pulses with higher photon energy than the present-day alternatives.

For the single-shot exposure, the required number of photons is ~$10^{12}$ to $10^{13}$ [139]. BISER driven by 20 TW lasers provides ~$10^{10}$ photons/pulse (experiment [23]), which scales to (expected) ~$10^{12}$ photons/pulse with a 200 TW driver [22]. Further, till now BISER was collected using small x-ray mirrors with a limited acceptance angle, thus, orders of magnitude larger number of photons is expected if the full BISER beam is collected.

Trains of attosecond pulses can be an indispensable tool for "videorecording" of ultra-fast processes in laser plasma, including surface dynamics. For applications requiring isolated attosecond pulses, methods similar to those used with atomic harmonics [143] can be applied.

A remarkable advantage of the BISER source is its compactness: the typical driving laser size is ~10-20 m in length, while the laser-plasma interaction region is sub-millimeter. We note that the laser and its beamline do not get radio-activated, thus the potentially radio-activated region has sizes of less than a meter. This is much smaller compared to XFELs (a ~kilometer-long radio-activated zone), and thus the BISER source is much safer for implementation in university and factory laboratories.

Another prospect of the BISER research is its application in laboratory astrophysics [30-35], where scaled experiments, or experiments with limited similarity [144] can be performed to test novel theories of singularity-based models of bright cosmic sources of radio waves (such as Fast Radio Bursts, FRBs [145]), gamma radiation (such as Gamma Ray Bursts, GRBs [146]), or yet-to-be-discovered gravitational BISER from singularities formed by usual [147] or dark matter flows [148].

These prospects show the urgency of the BISER source development, including the steps yet to be performed, such as the BISER pulse measurement and optimization – including isolated attosecond pulse generation, driving laser stabilization, full BISER beam collection using larger x-ray mirrors, experiments with multi-terawatt long-wavelength infrared lasers on the diagnostics of singularities and BISER radiation characterization, and testing – either experimentally or with simulations – the hypotheses, such as the proportionality of the source size to the emitted wavelength.

## 10. Conclusions

We have experimentally discovered a new mechanism of bright temporally- and spatially-coherent x-ray generation in laser plasma using high-quality multi-terawatt femtosecond lasers. We have proposed a new general effect, Burst Intensification by Singularity Emitting Radiation (BISER): an extreme enhancement, due to constructive interference, of the intensity of travelling waves from emitting matter extremely concentrated in converging flows. The extreme concentration of matter and BISER robustness is explained by catastrophe theory, which guarantees that the multi-stream flows lead to the formation of structurally-stable density singularities.

A synergy of experiment, theory and PIC simulations revealed the unique properties of the BISER sources in laser plasma: the perfect optical coherence (spatial and temporal) indicated by high-order harmonic comb spectra, attosecond-scale duration of pulses, and nano-scale source size. Consequently, BISER exhibits high brightness, by orders of magnitude exceeding that of third-generation synchrotrons and comparable to free electron lasers. Deduced scalings of BISER with the laser power and intensity promise reaching the keV photon energy range and terawatt-level attosecond x-ray pulses.

BISER paves the way to a compact bright coherent x-ray or XUV source for pumping, probing, imaging, or attosecond science, built on an in-lab or in-fab repetitive laser and a highly accessible, replenishable, and debris-free gas jet target.

We have proposed a method for imaging fast-moving (relativistic) singularities in laser plasma using BISER in the form of diffraction.

We have demonstrated that BISER is intrinsically very robust and that its reproducibility hinges on the laser stability. The current consensus is that there is still much potential for improving the control and stability of high-power laser systems before intrinsic physics limitations prohibit further improvement. Moreover, the trends revealed in our work suggest that higher laser power and access to different wavelengths will translate into a more powerful and versatile BISER. Further, higher laser repetition rates would be very useful. The many attractive applications of BISER will thus directly benefit from investments into the quality and versatility of high-power lasers.

We have predicted that, due to the high pulse energy and short duration, the x-ray BISER source in laser plasma has a potential to portray, for the first time, quantum systems on their proper, atomic timescales (~24 attoseconds).

We further note that advanced BISER experiments worth dedicated setups: in particular, the relativistic singularity measurements (Section 5), attosecond pulse duration measurements (Section 7), and full-beam BISER collection and quantum imaging with BISER x-ray source (Section 9) demand a dedicated BISER beamline and experimental area rather than the few-week beamtime offered at general user facilities, where the entire setup should be created from scratch every time.

Surprisingly, the nanometer size and attosecond duration of the BISER source in laser plasma are orders of magnitude smaller than, respectively, the minimum characteristic scales (wavelengths and oscillation periods) of plasma or laser. Although the laser-to-BISER energy transformation is relatively small, the combination of a short duration and small source size produces very high power and brightness. A similar property can have BISER in general, for different types of medium and travelling waves. In particular, the BISER concept creates a new framework for interpreting astrophysical observations, where media emitting electromagnetic and/or gravitational waves exhibit multi-stream flows and rapidly-moving density singularities, which can be the progenitors of gamma-ray bursts, fast radio bursts, and a new class of waveforms in modern gravitational wave astronomy [149].

## 11. Acknowledgments

We acknowledge fruitful and inspiring discussions with Drs. Igor Pogorelsky and Nicholas Dover. The research is supported by JSPS Kakenhi JP23H01151 and JP25H00621 and MEXT JPMXS0450300221.

## FIGURES

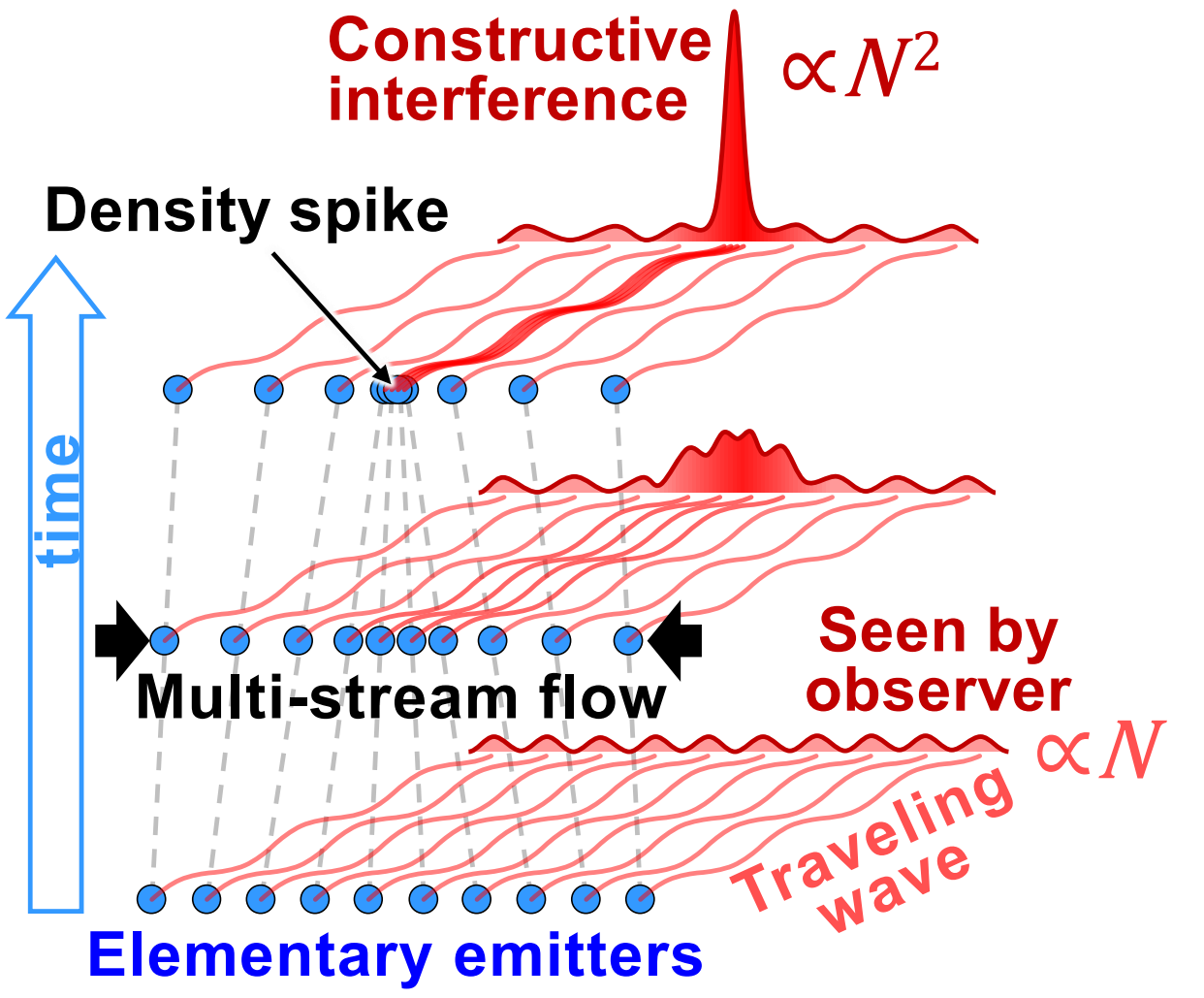

**Fig. 1.** Burst Intensification by Singularity Emitting Radiation (BISER) mechanism [23]. A converging flow of elementary emitters forms a density spike, where constructive interference occurs, thus causing a bright burst of temporally and spatially coherent radiation of travelling waves. Spatial distribution of emitters in the horizontal direction is shown in 3 moments of time; the vertical direction corresponds to time. The emitted waves travel into the plane of the figure.

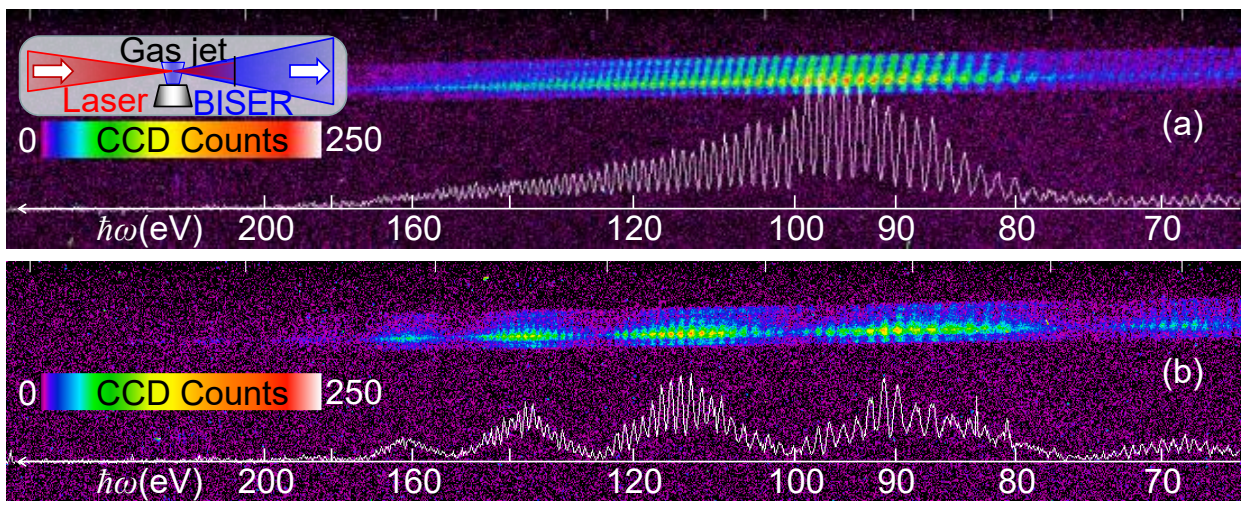

**Fig. 2.** Examples of single-shot raw data from the experiment [22] with the 9 TW J-KAREN laser. The inset shows a simplified sketch of the setup.

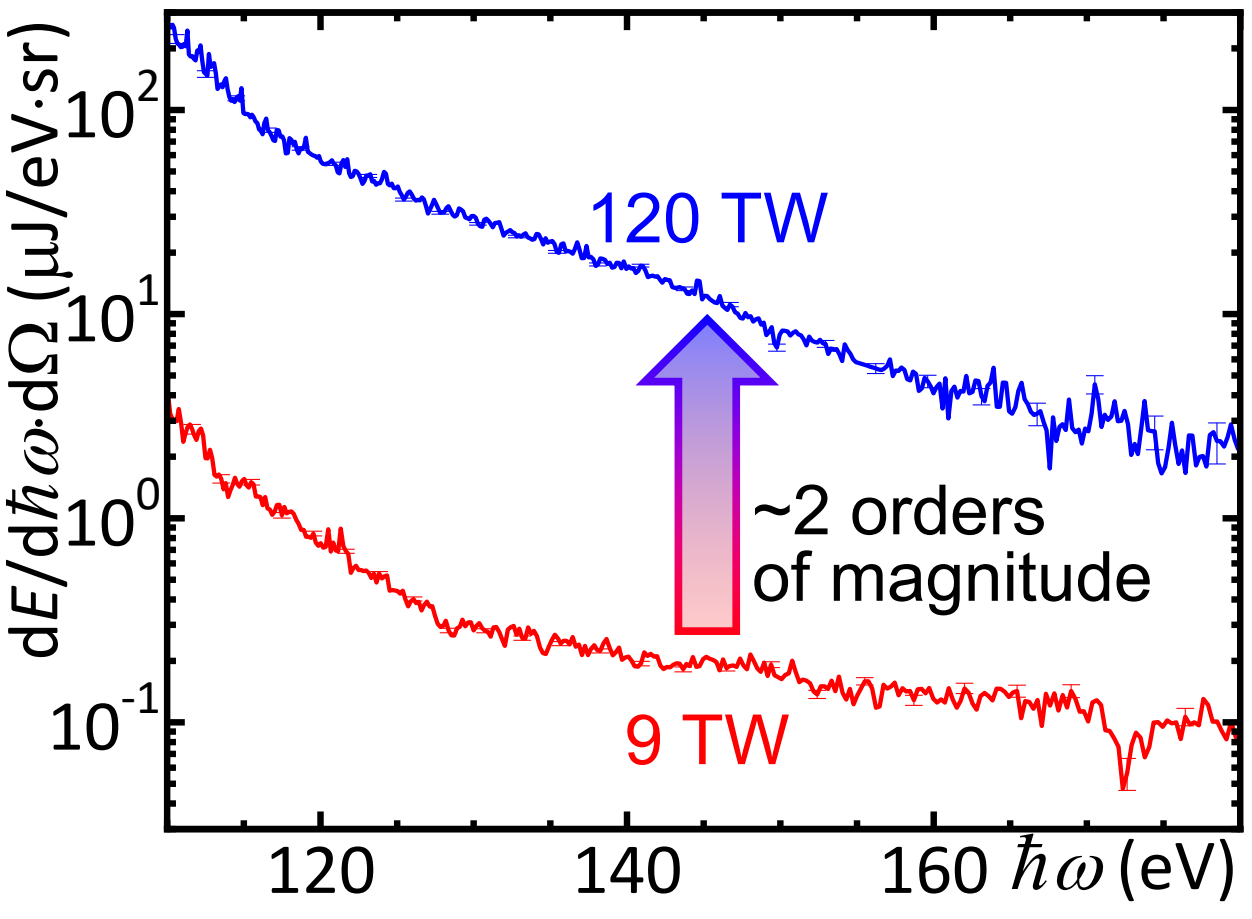

**Fig. 3.** BISER spectra driven by 9 TW and 120 TW lasers [48].

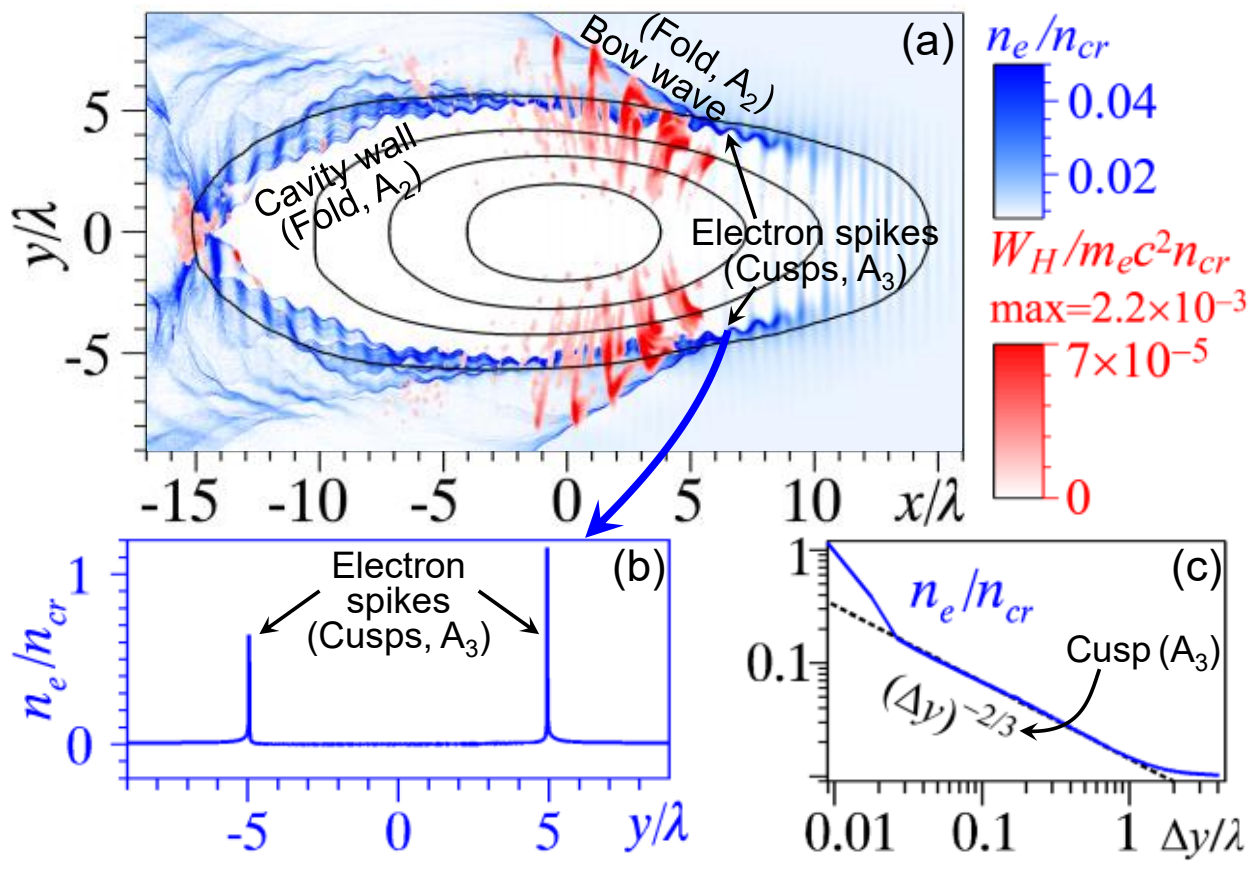

**Fig. 4.** BISER in PIC simulation [48]. (a) The electron density $n_e$, laser envelope (curves for $a = 1,4,7,10$), and electromagnetic energy density $W_H$, for frequencies from 60 to $100\omega$ at the same moment of time. (b) Lineout of the electron density at the position of the cusp, 10 laser cycles earlier than (a). (c) The right spike structure.

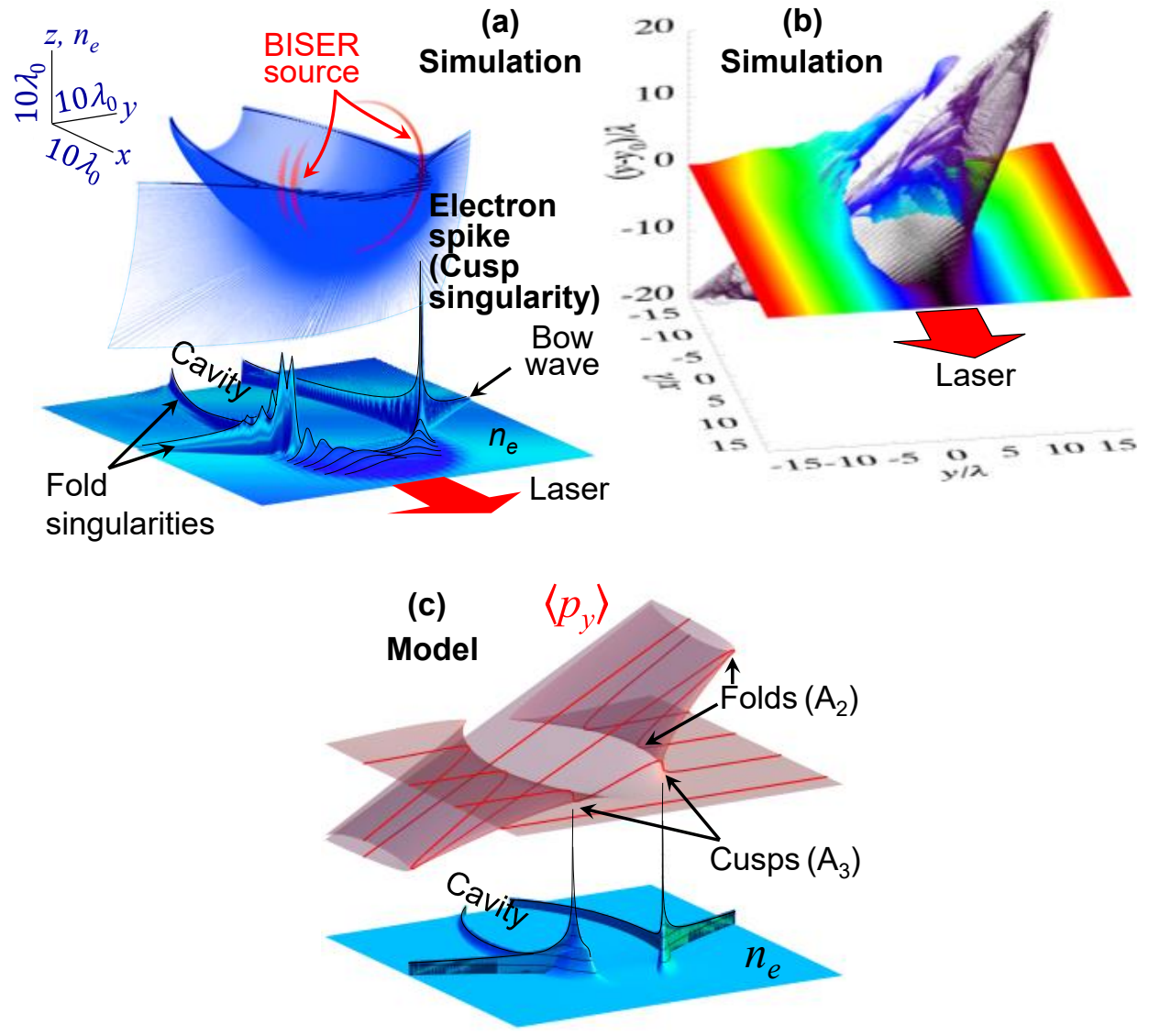

**Fig. 5.** 3D structure of the bow wave and the BISER source [48]. (a) 3D PIC simulation: electron density $n_e$ (top, upper half removed) and its cross-section at $z = 0$ (bottom). The electromagnetic energy density $W_H$ for frequencies $\geq 4\,\omega$ (appears as arcs). (b) Displacement with respect to the initial position along $y$ axis of electrons forming $n_e$. Before the driver pulse enters, electrons at rest fill the plane $(x, y, z = 0)$; the color scale represents the initial electron coordinate magnitude, $|y_0|$. (c) Catastrophe theory model: singularities in the electron density are created by foldings of the electron phase space $(x, y, \langle p_y \rangle)$; this correspondence of the phase space folds and density singularities is essentially the same as in (a).

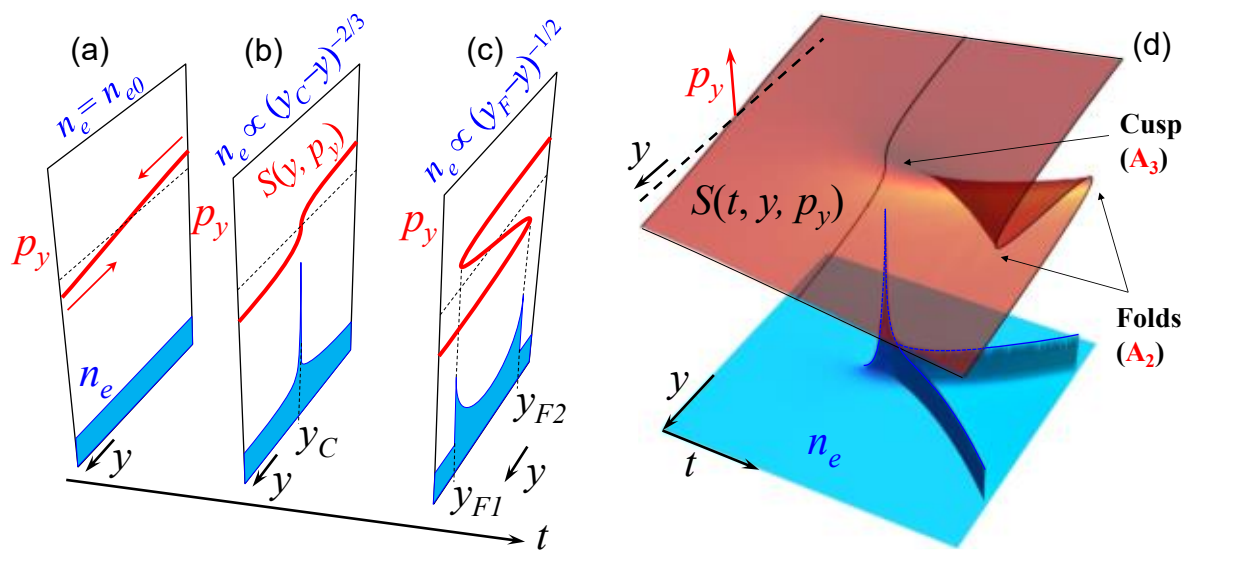

**Fig. 6.** Catastrophe theory model [48] of electron flow with spatially varying momentum, $p_y$. (a), (b), and (c) The evolution of the electron phase volume $S(y, p_y)$ at different moments of time $t$. (d) The extended phase volume of electrons $S(t, y, p_y)$, for which (a), (b), and (c) are lineouts for fixed $t$. The projection of $S(t, y, p_y)$ onto $y$ axis produces the electron density $n_e(t, y)$, according to Eq. (1). The initially homogeneous (a) distribution of electrons evolves into the cusp singularity (b), Eq. (5), that decays into two fold singularities (c), Eq. (4).

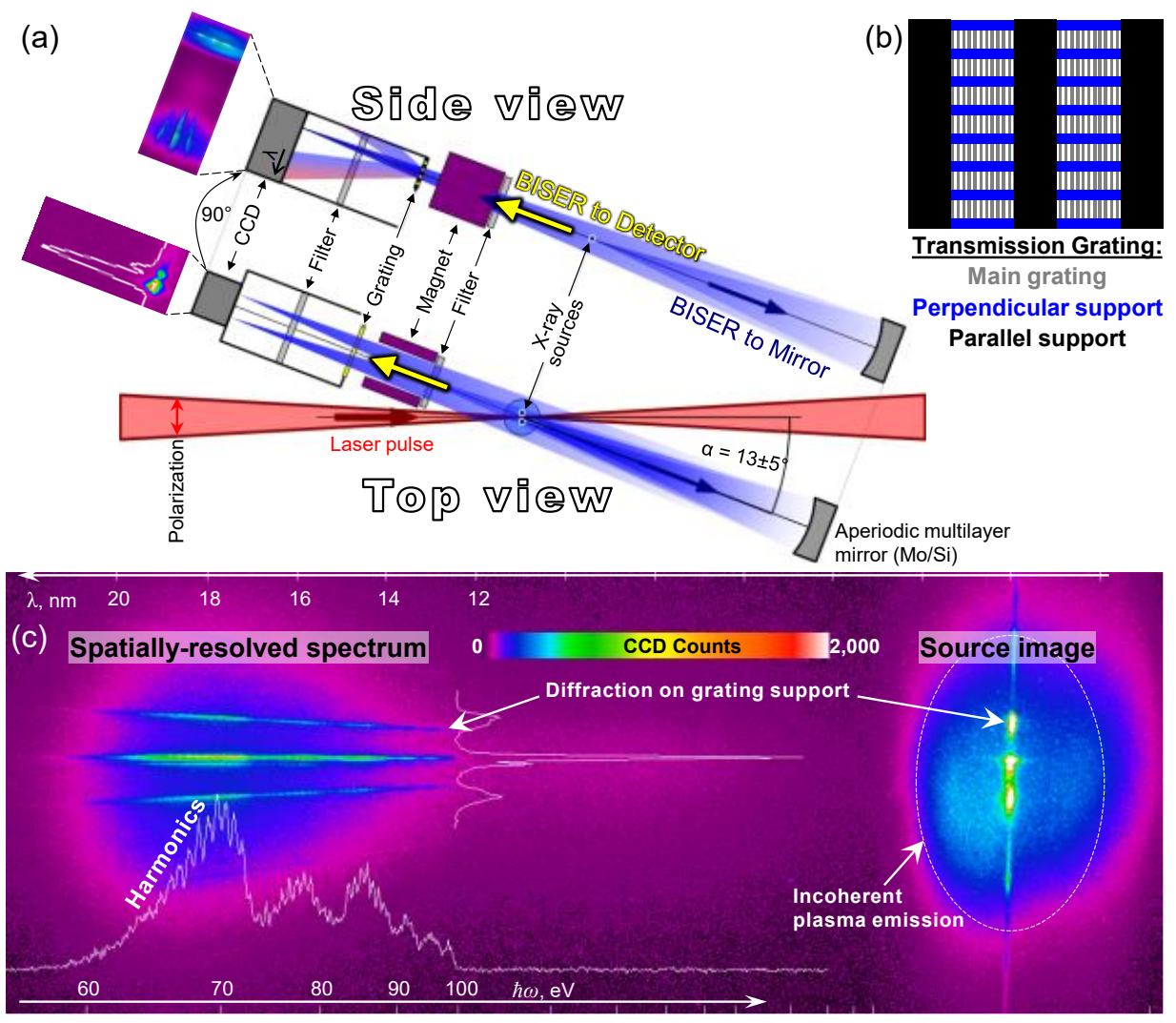

**Fig. 7.** Spatially-resolved BISER spectroscopy [23] showing double source for linearly polarized driving laser. (a) Setup sketch shows a side-view and top-view of the imaging XUV spectrograph. (b) Transmission grating schematic. (c) Single-shot raw data consisting of the 0th diffraction order (Source image) and the 1st diffraction order of the main grating (Spatially-resolved spectrum) and its lineouts (white lines). Both the 0th and the 1st diffraction orders are further diffracted by the grating perpendicular support in the vertical direction.

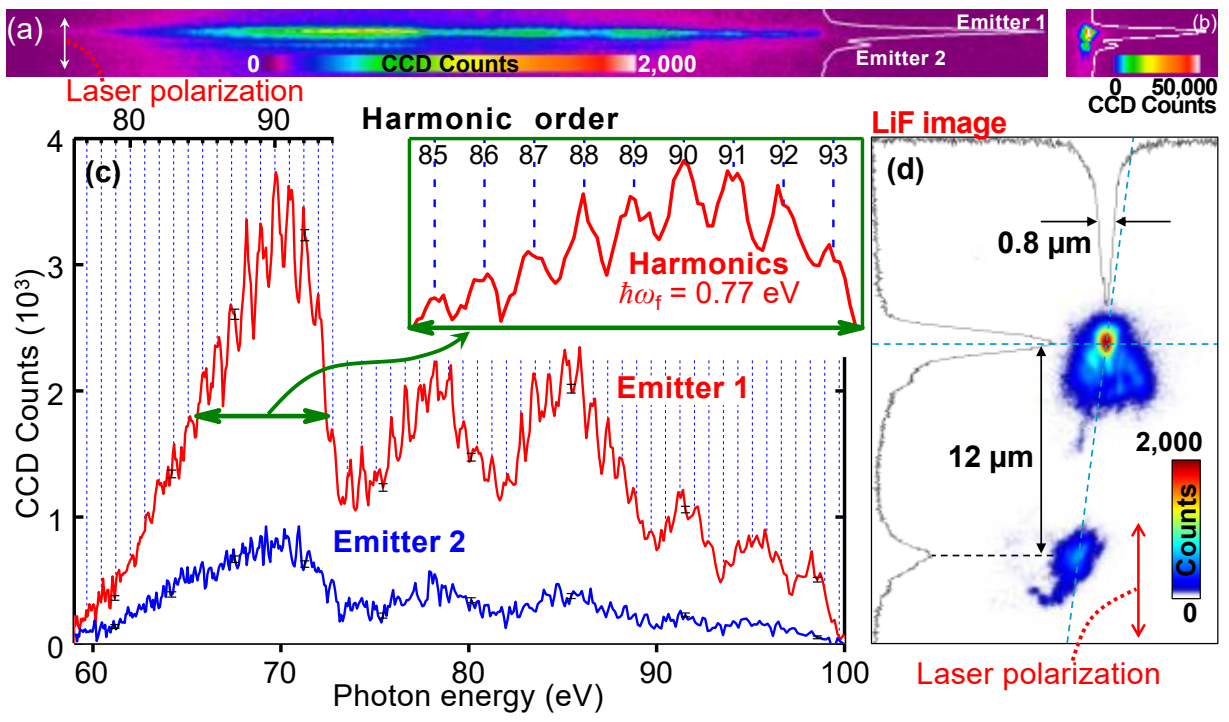

**Fig. 8.** Double BISER source, experiment [23]. (a) Magnified view of the spatially-resolved spectrum of Fig. 7(c). (b) Magnified view of the source image of Fig. 7(c); note the color scale difference compared to (a). (c) Lineouts of the spectra of emitters 1 and 2 in (a). (d) Single-shot image recorded with a high-resolution LiF detector showing a nano-scale double source. Emitters in (a-d) appear non-equal due to off-axis observation which preferentially detects the emitter 1.

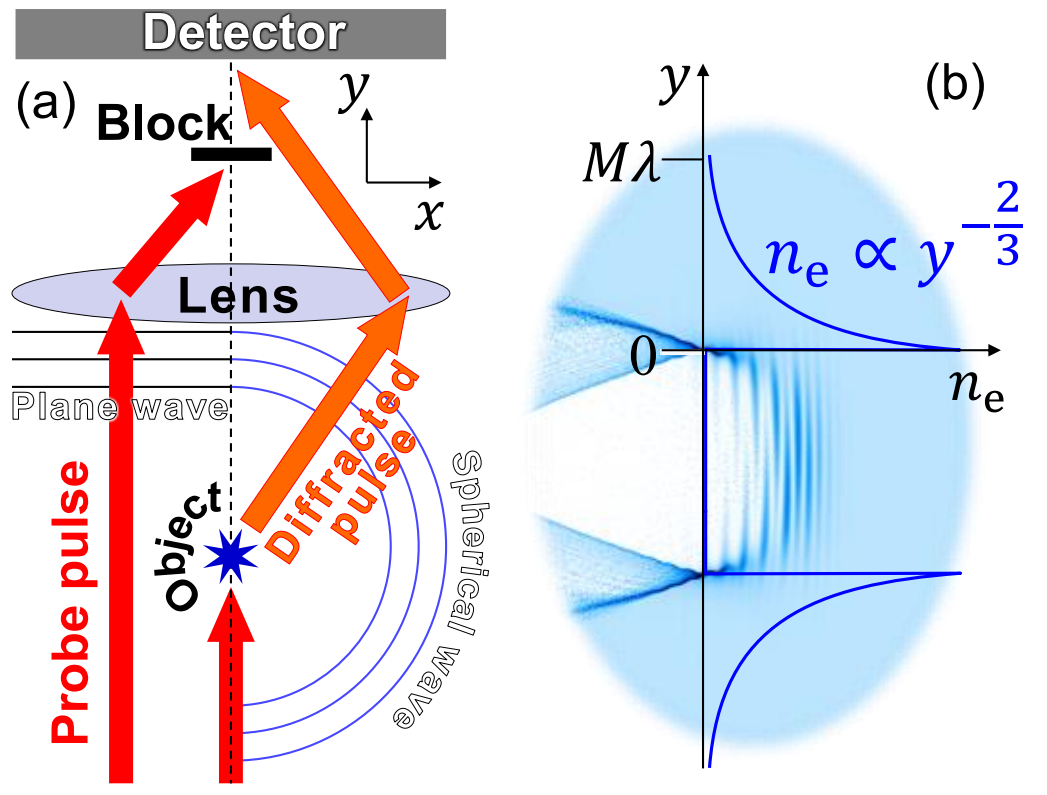

**Fig. 9.** Schlieren imaging of density singularities [101]. (a) Rays and wave fronts transmitted through the lens, assuming rotational symmetry with respect to the dashed line. On the left: the probe pulse is blocked by an opaque filter. On the right: the diffracted radiation is (partially) focused onto the detector, forming the object image. (b) The scheme for the refractive index integration (see the text).

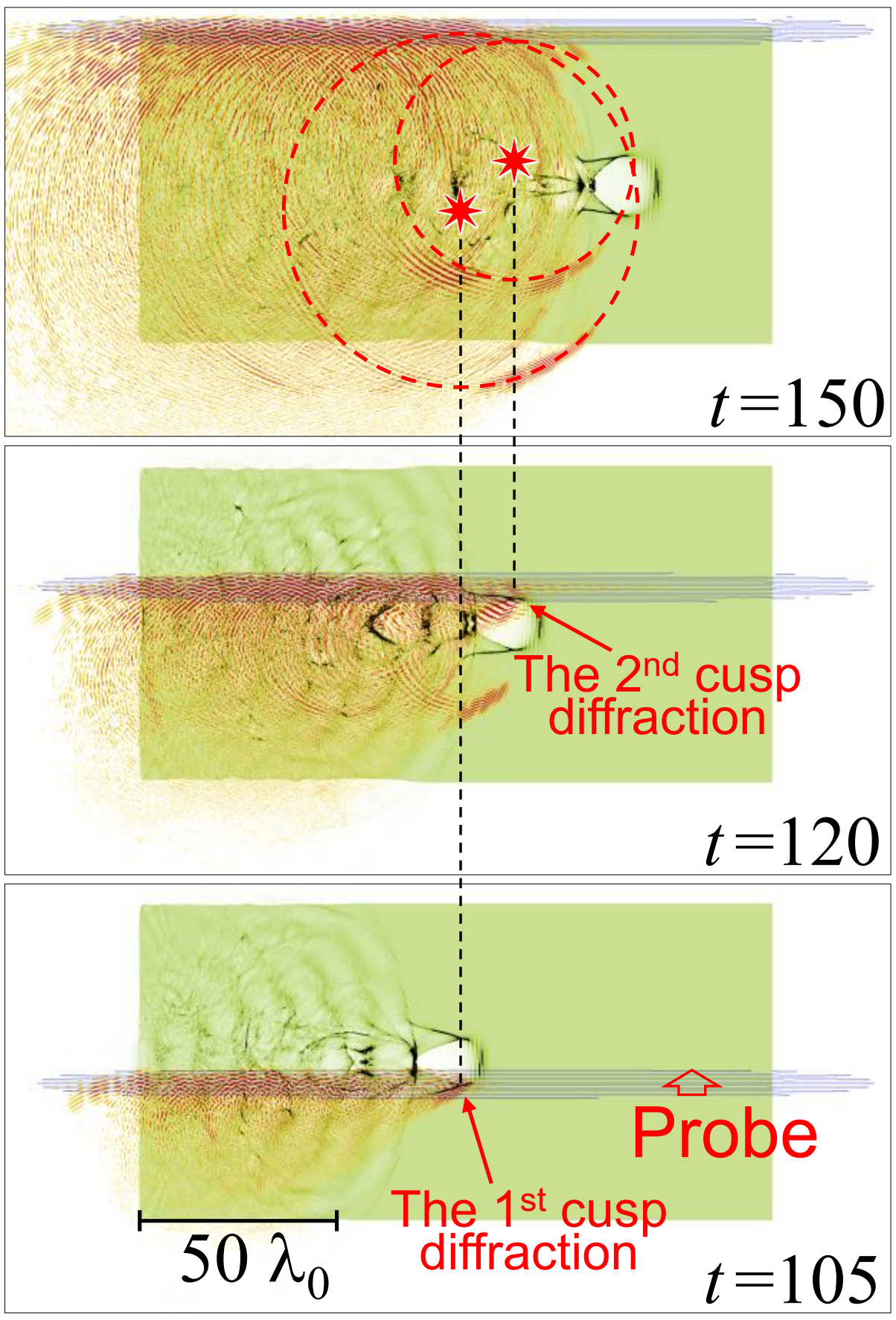

**Fig. 10.** PIC simulation of the probe diffraction on the singularities created by the driver [101]. The Driver laser pulse is revealed by density modulations at the first wake wave cavity front. The probe pulse corresponds to nearly horizontal curves for $E_z = 10^{-3}$. The diffracted radiation corresponds to $0 < E_z < 5 \times 10^{-4}$(white-red color scale). Black curves for $n_e = 3n_{e0}$; white-green color scale for $0 < n_e < 8n_{e0}$.

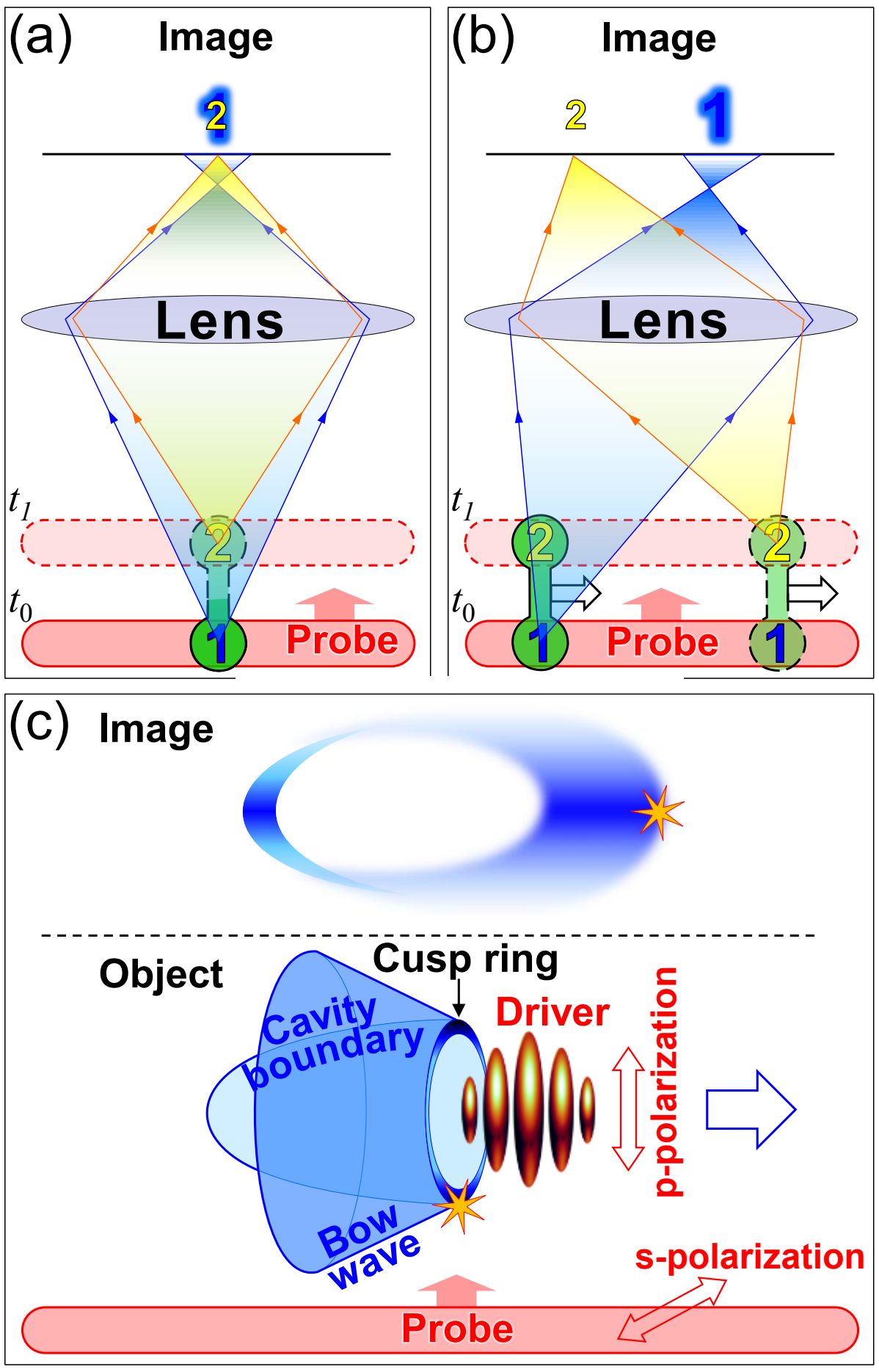

**Fig. 11.** Effects in imaging of moving constellation of singularities [101]. A short probe laser pulse irradiates a static (a) and moving (b) elongated object, one end at time $t_0$ and another end at time $t_1 > t_0$. (c) A sketch of an expected image of the cusp singularity in the experiment. The star shows which point of the object corresponds to which point in the image. As in frames (a) and (b), the point nearest to the detector of the cusp ring is at the focus. It appears as the leftmost spot of the ellipse in the image.

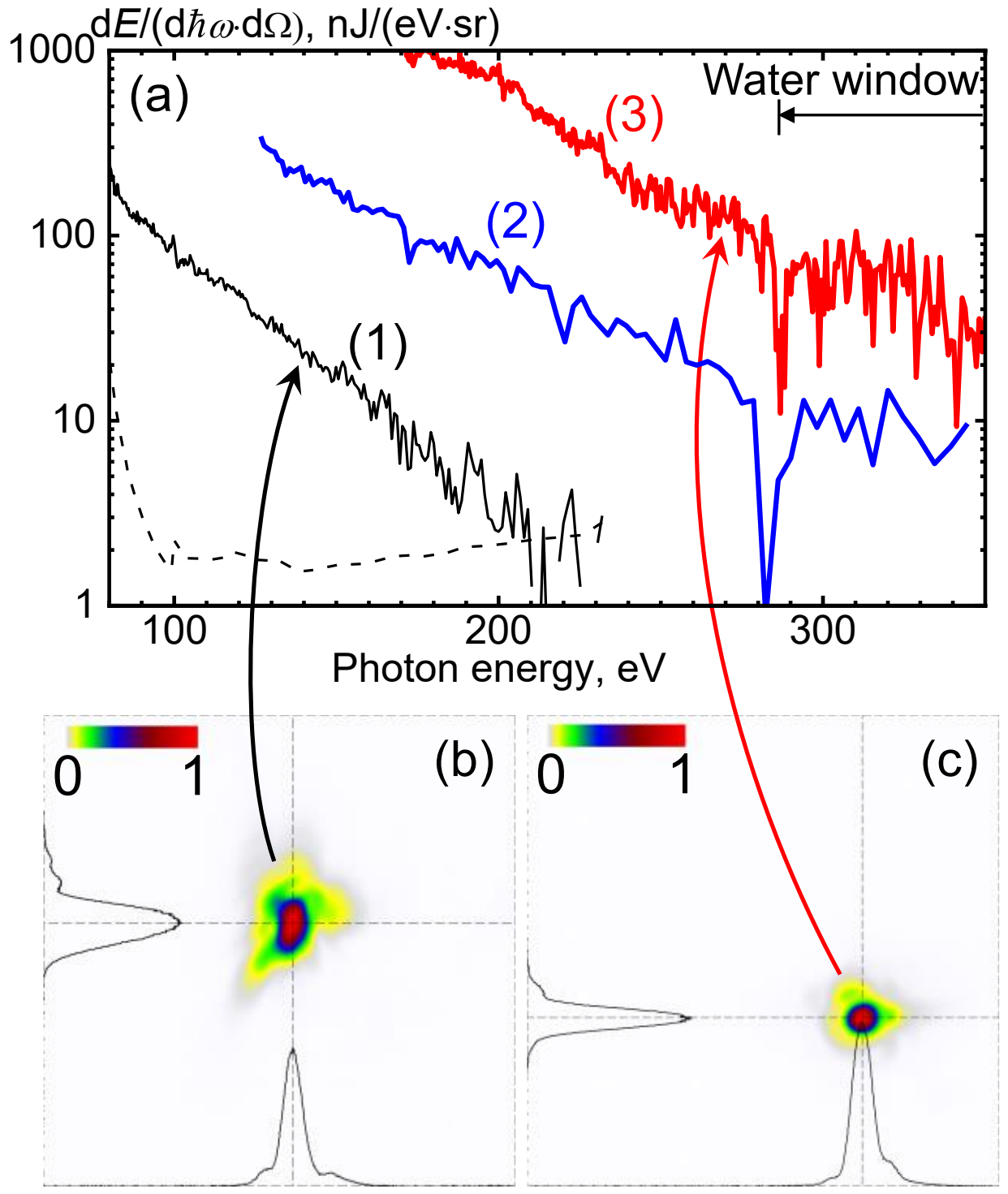

**Fig. 12.** Influence of the focal spot quality on BISER [108]. (a) BISER spectra obtained under similar experimental conditions but with different focal spot quality: curves (1) and (3) from the experiments [108], curve (2) from the experiment [22]; the dashed line is the noise level for (1). (b,c) Focal spot shapes corresponding to curves (1) and (3) in (a).

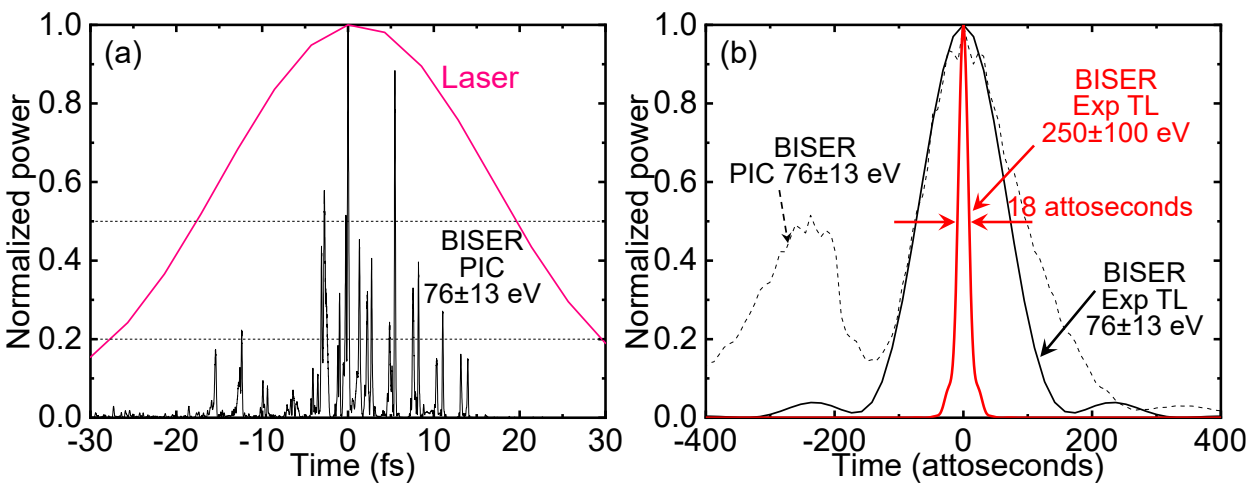

**Fig. 13.** BISER temporal shape. (a) 2D PIC simulation [23] showing attosecond pulse train in the 76±13 eV spectral range. There are 4 pulses above the 50% intensity level and 14 pulses above the 25% intensity level. (b) Dashed line, the main pulse from 2D PIC simulation shown in (a); Black line, the transform-limited pulse calculated from the experimental spectrum [23] in the same spectral range; Red line, the 18-attosecond transform-limited pulse calculated from the experimental spectrum [108] shown in Fig. 12(a) by the red curve (3).

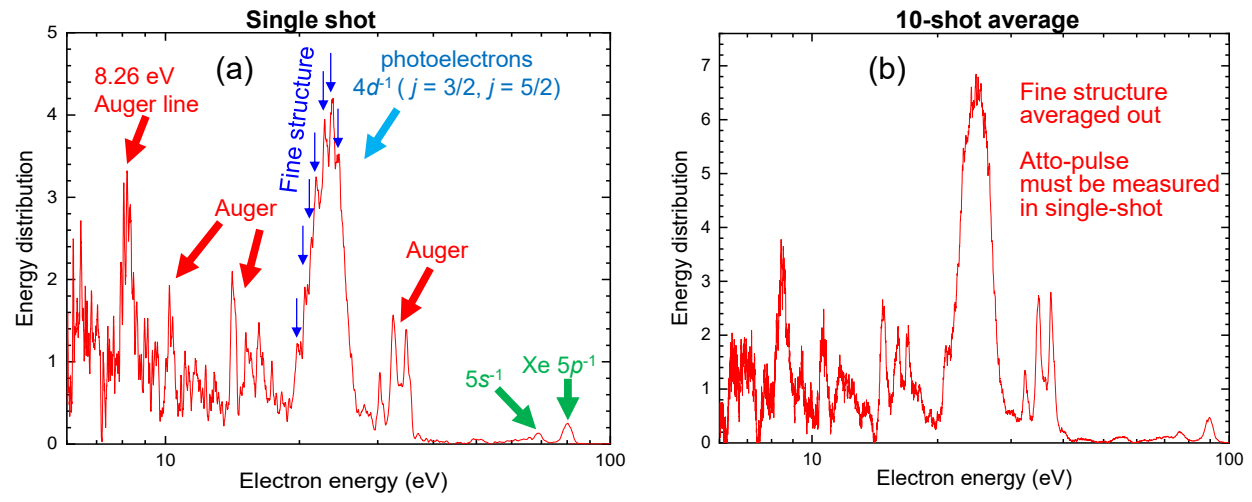

**Fig. 14.** Electron energy spectra obtained with the Time-of-Flight (ToF) Magnetic Bottle Electron Spectrometer (MBES) [123]. The shown energy range includes both photo- and Auger electrons produced by BISER in a secondary Xe gas target. (a) Single shot, note the fine structure on the photoelectron peak. (b) 10-shot average, note the absence of the fine structure.

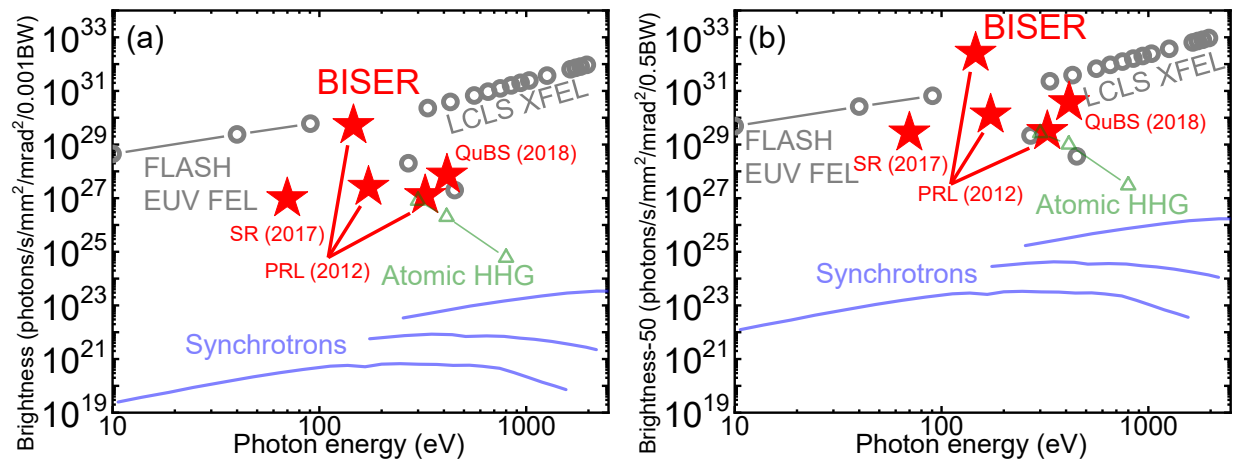

**Fig. 15.** Comparison of estimated peak spectral brightness of several x-ray sources. (a) Standard definition including number of photons in 0.1% bandwidth, suitable for narrowband sources like synchrotrons after monochromatization or femtosecond XFELs. (b) Definition with 50% bandwidth suitable for broadband sources, such as any attosecond x-ray source capable of ultimately short durations down to a few cycles of the x-ray wave.

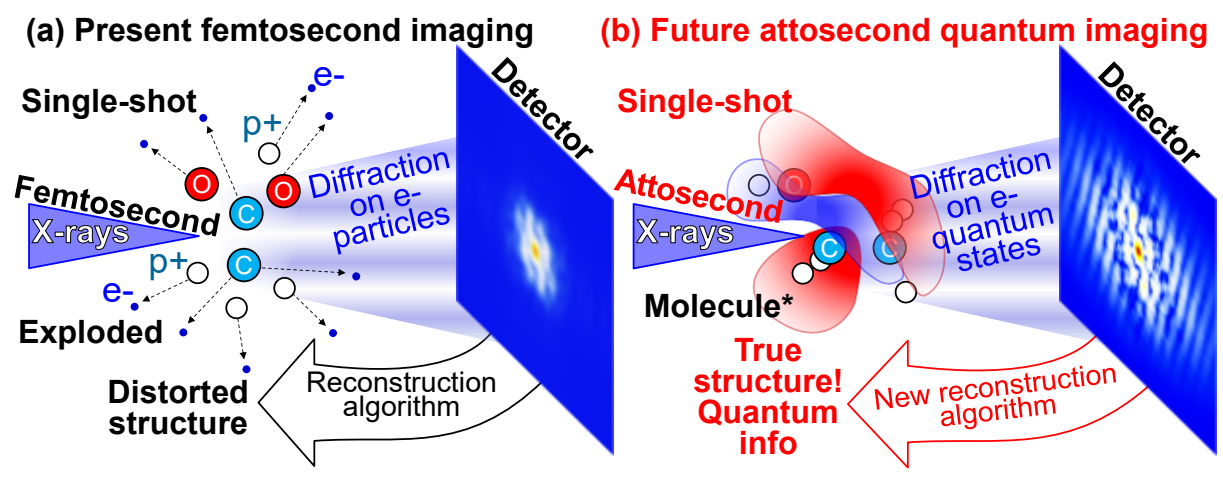

**Fig. 16.** Coherent Diffractive Imaging utilizing the Image-Before-Destroy concept [139]. (a) Present scheme with bright femtosecond pulses: light atoms are captured with distortion, while electrons and protons escape. (b) Prospective imaging of the quantum state with bright (capable of single-shot exposure) x-ray pulses shorter than the atomic time unit (24 attoseconds). *The artistic image of a $C_2H_5OH$ molecule is inspired by [150].

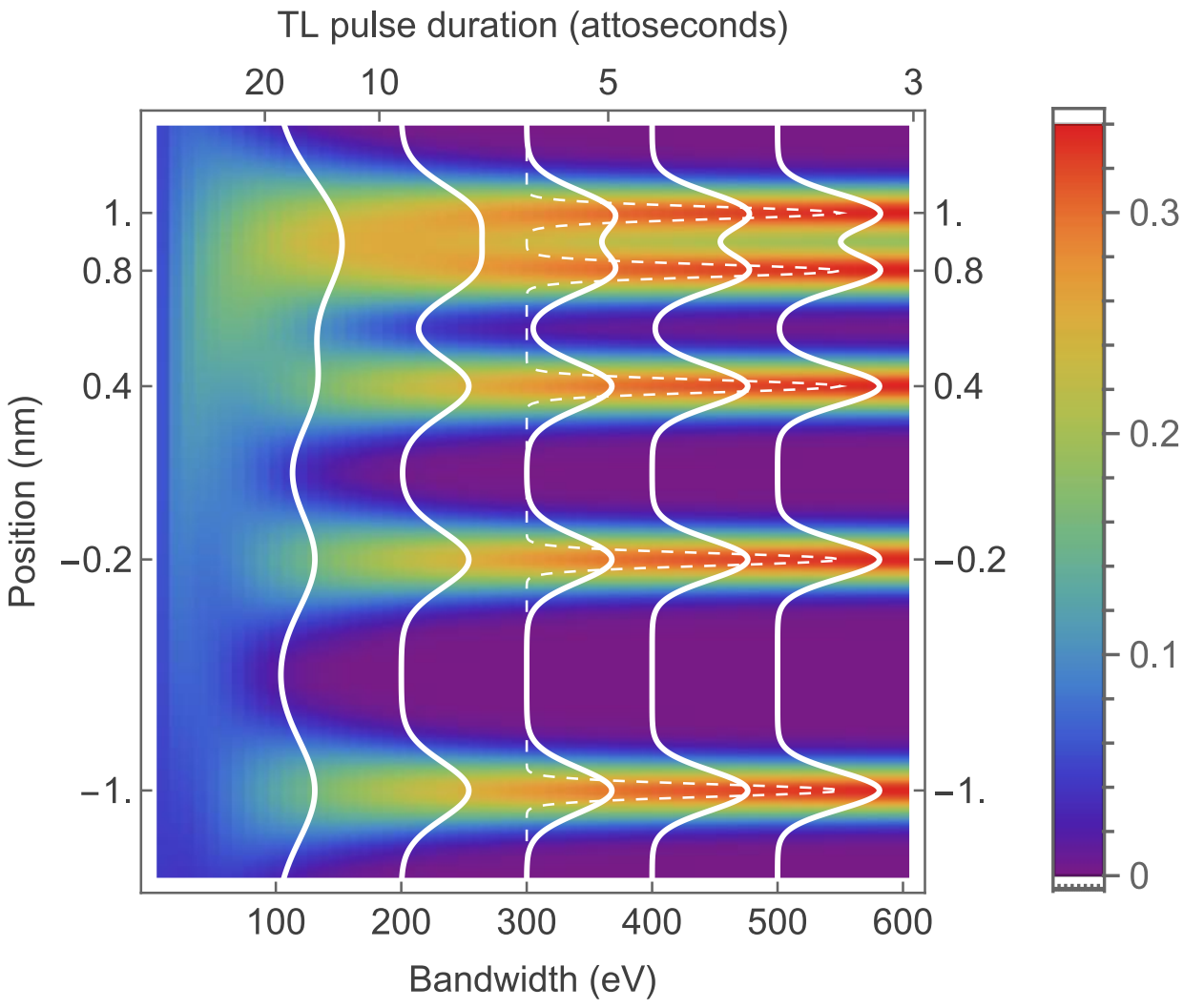

**Fig. 17.** Quasiclassically estimated degradation of resolution due to photoelectron motion blur assuming transform-limited Gaussian pulses with a central photon energy of 5 keV. The initial object has sharp features at the shown positions separated by 0.2, 0.4, 0.6, and 0.8 nm (white dashed line). The color scale shows object modulation blurred by the motion (the initial modulation amplitude is 1.0), with the white solid lines corresponding to 100, 200, 300, 400, and 500 eV bandwidths.